\documentclass[aps,prx,twocolumn,showpacs,preprintnumbers,amsmath,amssymb,superscriptaddress]{revtex4-1}

\usepackage{graphicx}
\usepackage{dcolumn}
\usepackage{color}
\usepackage{amsmath}
\usepackage[breaklinks,colorlinks,bookmarks=false,citecolor=blue,linkcolor=red,urlcolor=blue]{hyperref}

\begin{document}

%%%% Remove for preprint
\onecolumngrid
\vspace*{-12.5mm}

\rightline{Phys. Rev. B {\bfseries 93}, 054408 (2016)}
\vspace*{3mm}

\twocolumngrid
%\advance\textheight by 2 true mm
%%%%

\title{Thermodynamic properties of highly frustrated quantum spin ladders: \\
Influence of many-particle bound states}

\author{A. Honecker}
\affiliation{Laboratoire de Physique Th\'eorique et Mod\'elisation, CNRS 
UMR 8089, Universit\'e de Cergy-Pontoise, F-95302 Cergy-Pontoise Cedex, France}
\affiliation{Institut f\"ur Theoretische Physik, Georg-August-Universit\"at 
G\"ottingen, 37077 G\"ottingen, Germany}

\author{S. Wessel}
\affiliation{Institut f\"ur Theoretische Festk\"orperphysik, JARA-FIT and 
JARA-HPC, RWTH Aachen University, 52056 Aachen, Germany}

\author{R. Kerkdyk}
\affiliation{Institut f\"ur Theoretische Physik, Georg-August-Universit\"at 
G\"ottingen, 37077 G\"ottingen, Germany}

\author{T. Pruschke}
\thanks{Deceased.}
\affiliation{Institut f\"ur Theoretische Physik, Georg-August-Universit\"at 
G\"ottingen, 37077 G\"ottingen, Germany}

\author{F. Mila}
\affiliation{Institute of Physics, Ecole Polytechnique F\'ed\'erale
Lausanne (EPFL), CH-1015 Lausanne, Switzerland}

\author{B. Normand}
\affiliation{Department of Physics, Renmin University of China, Beijing
100872, P.~R.~China}

\date{November 4, 2015; revised January 6, 2016; published February 8, 2016} % \today}

\begin{abstract}
Quantum antiferromagnets have proven to be some of the cleanest realizations 
available for theoretical, numerical, and experimental studies of quantum 
fluctuation effects. At finite temperatures, however, the additional effects 
of thermal fluctuations in the restricted phase space of a low-dimensional 
system have received much less attention, particularly the situation in 
frustrated quantum magnets, where the excitations may be complex collective 
(bound or even fractionalized) modes. We investigate this problem by studying 
the thermodynamic properties of the frustrated two-leg $S = 1/2$ spin ladder, 
with particular emphasis on the fully frustrated case. We present numerical 
results for the magnetic specific heat and susceptibility, obtained from 
exact diagonalization and quantum Monte Carlo studies, which we show can be 
rendered free of the sign problem even in a strongly frustrated system and 
which allow us to reach unprecedented sizes of $L = 200$ ladder rungs. We 
find that frustration effects cause an unconventional evolution of the 
thermodynamic response across the full parameter regime of the model. 
However, close to the first-order transition they cause a highly anomalous 
reduction in temperature scales with no concomitant changes in the gap; 
the specific heat shows a very narrow peak at very low energies and the 
susceptibility rises abruptly at extremely low temperatures. Unusually, the 
two quantities have different gaps over an extended region of the parameter 
space. We demonstrate that these results reflect the presence of large 
numbers of multi-particle bound-state excitations, whose energies fall 
below the one-triplon gap in the transition region. 
\end{abstract}


\maketitle

\section{Introduction}
\label{sec:intro}

Quantum magnets rank among the simplest model systems in condensed matter 
physics, but nonetheless exhibit some of its most fundamental phenomena, 
including quantum phase transitions, symmetry-breaking and restoration, 
collective and fractionalized excitations, topological order, and complex 
entanglement of the quantum wave function. From a theoretical point of view, 
these phenomena arise even in models with only the most elementary magnetic 
interaction, described by the Heisenberg model \cite{Heisenberg}, and depend 
crucially on the geometry of the lattice. In low-dimensional systems, quantum 
fluctuation effects are strong, and models in one spatial dimension provide 
particularly good examples of exotic phenomena \cite{MiKo04}, not least the 
gapped ``Haldane'' state of the nearest-neighbor spin-1 chain \cite{haldane83} 
and the exactly dimerized ground state, with fractional ``spinon'' excitations,
of the frustrated spin-1/2 $J_1$-$J_2$ chain \cite{MG69,Ma69,ShaSu81j1j2}. In higher 
dimensions, models on bipartite lattices tend to show semi-classical 
long-ranged magnetic order, whereas geometrically frustrated lattices offer 
both analogs of the unconventional ground states of quantum spin chains and 
uniquely high-dimensional phases such as the quantum spin liquid
\cite{RSH04,BalentsRev,Fbook,DiepBook}.

A frustrated two-dimensional system related to the $J_1$-$J_2$ chain is 
the Shastry-Sutherland model \cite{ShaSu81}, which exhibits an exact  
dimer-singlet ground state. The compound SrCu$_2$(BO$_3$)$_2$, based on 
$S = 1/2$ Cu$^{2+}$ ions, provides not only a good realization of this model, 
but one that is believed to be located close to a frustration-induced quantum 
phase transition out of the dimer-singlet phase \cite{MiUe03,Kageyama99,
Kageyama00a,Kageyama00b}. As a consequence, SrCu$_2$(BO$_3$)$_2$ displays 
a number of exotic properties, including a magnetization curve that exhibits 
many plateaus \cite{Kageyama99,Kageyama99b,Kodama02,Jaime12,Matsuda13}. The 
origin of this unusual behavior lies in the strong suppression of kinetic 
energy contributions from the triplet excitations, due to the almost 
perfectly frustrated coupling of the dimers in the Shastry-Sutherland 
model \cite{MoTo00}. When magnetic excitations become highly localized in 
this way, the related phenomenon of bound-state formation appears. In 
two-leg spin-1/2 ladders with little or no frustration, bound states are 
present in the spectrum at higher energies \cite{ZHSTM01,Windt01}, whereas 
in strongly frustrated systems they can occur at low energies, particularly 
close to a quantum phase transition as in the Shastry-Sutherland model 
\cite{MoTo00,KBHU00}. 

The presence of additional low-energy states in the spectrum of a system 
is expected to have a clear signature in its physical properties. Despite 
the fact that the thermodynamic properties are frequently used as one of 
the first experimental characterizations of any new material, their 
computation actually constitutes a major challenge for theory. Exact 
results for the finite-temperature properties of truly quantum models are 
scarce, being restricted to integrable systems such as the nearest-neighbor 
spin-1/2 Heisenberg chain \cite{Takahashi99,Gaudin71,EAT94,Kl98,KlJ00,
THKO10}, which can be solved by generalizations of the Bethe Ansatz
\cite{Bethe31,BaxterBook}.

Among the numerical approaches to this problem, one of the first to be used 
was the full exact diagonalization (ED) of the Hamiltonian in the exponentially 
large Hilbert space \cite{rbf}. Although this method can be extended as far 
as 24 spin-1/2 sites with current technology \cite{HMHV06}, standard 
applications are limited to approximately 20 $S = 1/2$ sites \cite{MiUe03}, 
a size with restricted applicability in two and higher dimensions. Another 
classic approach is the use of high-temperature series expansions 
\cite{DoGr74,LSR11,LSR14}, but frustrated systems such as the 
Shastry-Sutherland model \cite{WHO99} illustrate the difficulties suffered 
by this method in accessing the low-temperature regime. Finite-temperature 
variants of the density-matrix renormalization group (DMRG) technique 
\cite{WhiteDMRG} such as a quantum transfer-matrix formulation \cite{WangXiang97,Xiang98},
imaginary time evolution in an enlarged Hilbert space \cite{FW05},
and minimally entangled typical thermal states \cite{White09}
have been developed into powerful tools for calculating 
the thermodynamic properties of one-dimensional systems,
but despite the development of modern matrix-product-state 
formulations \cite{Schollwock11}, these methods 
have mostly remained limited to one dimension. Quantum Monte Carlo (QMC) 
simulations provide the most flexible approach in higher dimensions, and are 
able to resolve the primary features of the thermodynamic quantities for models 
that are at most weakly frustrated \cite{Johnston00b,cav2o4,Johnston11}, but 
these break down in more strongly frustrated systems due to the notorious QMC 
sign problem.

\begin{figure}[t]
\centering\includegraphics[width=0.8\columnwidth]{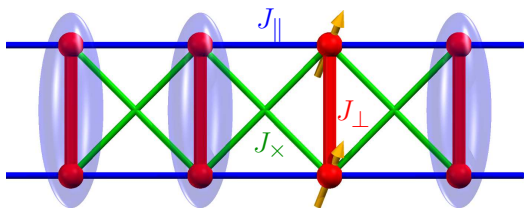}
\caption{(Color online) Representation of superexchange interactions in 
a frustrated spin ladder. Each ladder site (spheres) hosts an $S = 1/2$ 
quantum spin and the Heisenberg couplings between spins are specified by the 
parameters $J_\perp$ for the ladder rungs, $J_\|$ for the ladder legs, and 
$J_\times$ for the cross-plaquette couplings, which we take to be symmetrical. 
Ellipsoids represent singlet states of the two rung spins and their absence 
a rung triplet.}
\label{fig:s}
\end{figure}

As a consequence of these technical limitations, there has been little 
systematic investigation of the finite-temperature response of quantum spin 
systems with strong geometrical frustration, and thus the influence both of 
bound states in the excitation spectrum and of proximity to a quantum critical 
point on the thermodynamic properties remains poorly understood. Here we 
provide a detailed analysis of these issues using the model of the fully 
frustrated two-leg spin-1/2 ladder, which is shown in Fig.~\ref{fig:s}. A 
significant amount of information is already available concerning the 
zero-temperature properties of this precise model \cite{gelfand91,BoGa93,
rx,Weihong98,KSE99,Wang00,rhmt,lamas15} and variants thereof \cite{rbb,MiTo02,
Starykh04,rvh,Kim08,HiSta10,Poilblanc10,CCBZ15}. Like the Shastry-Sutherland 
model, it exhibits a phase with a dimer-singlet ground state, fully localized 
triplet excitations, and a first-order phase transition \cite{gelfand91,rx}.
The importance of multi-triplet bound-state excitations was also recognized 
in early analytical studies of the fully frustrated ladder \cite{gelfand91,rx} 
and observed in a DMRG calculation \cite{Wang00}. 

Building on these analytical properties, we perform and interpret detailed 
numerical calculations of the magnetic specific heat and susceptibility of 
the fully frustrated ladder. With a view to a complete microscopic 
understanding of the spectrum, and profiting from the very short 
correlation length of the fully frustrated system, our tool of choice for 
this investigation is ED rather than a finite-temperature variant of DMRG. 
We will show that this model permits a detailed analysis of the excitation 
spectrum, which shows a highly unconventional emergence of multi-particle 
bound states involving very many rungs near the quantum phase transition. 
Because this feature exceeds the system-size limitations of ED, we exploit 
the properties of the model to express the spin Hamiltonian in the dimer 
basis and achieve the ability to perform QMC simulations completely free 
of the sign problem. These we employ to compute the numerically exact 
specific heat and susceptibility for ladders of up to 400 spins, which we 
show is well in the thermodynamic limit even for systems arbitrarily close 
to the quantum phase transition.

The structure of this article is as follows. In Sec.~\ref{sec:ffl} we 
introduce the fully frustrated ladder model and highlight some of its 
analytical properties. These allow us to review the ground-state phase 
diagram, to discuss the nature of the low-energy excitations in the two 
different phases, which include exact and perfectly localized bound states,
and to highlight the appearance of many low-lying states near the phase
transition. We then use these properties in an analytical discussion of a 
simple approximation to the thermodynamics of the fully frustrated ladder 
in the rung-singlet phase and compare its qualitative features with those 
of unfrustrated ladders. Because a systematic analysis of thermodynamic  
properties, particularly near the phase transition, requires exact numerical 
results, in Sec.~\ref{sec:methods} we explain how the specific properties of 
the fully frustrated ladder may be exploited to perform maximally efficient 
ED calculations and, despite the frustration, sign-problem-free QMC 
simulations. Section \ref{sec:results} presents and compares the results we 
obtain from these numerical calculations for the magnetic specific heat and 
susceptibility over the full range of temperatures, for a selection of fully 
frustrated ladders with different coupling ratios. In Sec.~\ref{sec:interp} 
we provide some analytical interpretations of our results, by comparing them 
with small- and large-cluster approximations, and by discussing the different 
energy scales characterizing the dramatic effects occurring near the phase 
transition. Our conclusions concerning the key role of low-lying multi-triplet 
bound states in determining the thermodynamic properties are summarized 
briefly in Sec.~\ref{sec:summary}.

\section{Fully Frustrated Ladder}
\label{sec:ffl}

The frustrated Heisenberg spin ladder we consider is represented schematically 
in Fig.~\ref{fig:s}. In addition to the ``rung'' interaction, $J_\perp$, 
defining the fundamental dimer unit, and the ``leg'' interaction, $J_\|$, 
defining the two chains, we include a symmetrical cross-plaquette coupling, 
$J_\times$, which frustrates $J_\|$. The Hamiltonian of the model for any spin 
quantum number, $S$, and for a ladder of $L$ rungs is  
\begin{equation} 
H = \sum_{i} \! J_\perp {\vec S}_{i}^1 \cdot {\vec S}_{i}^2 + \!\! \sum_{i,m=1,2} 
\!\! (J_\| {\vec S}_{i}^m \cdot {\vec S}_{i+1}^m + J_\times {\vec S}_{i}^m \cdot 
{\vec S}_{i+1}^{\bar m}) ,
\label{eq:essh} 
\end{equation} 
where $i$ is the rung index, $m = 1$ and $2$ denote the two chains of the 
ladder, and ${\bar m}$ is the chain opposite to $m$. In our numerical 
calculations we will impose periodic boundary conditions, such that $i + L 
\equiv i$.

The primary focus of our investigation is the fully frustrated case, 
$J_\times = J_\|$. In this situation, the Hamiltonian (\ref{eq:essh}) can be 
reexpressed in the form \cite{rx,rhmt}
\begin{equation} 
H = J_\| \sum_{i=1}^L \vec{T}_i \cdot \vec{T}_{i+1} + J_\perp \sum_{i=1}^L 
\left(\frac{1}{2} \, \vec{T}_i^2 - S\,(S + 1)\right) \! ,
\label{eq:exeh} 
\end{equation}
where $\vec{T}_i = \vec{S}_{i}^1 + {\vec S}_{i}^2$ is the total spin of 
rung $i$ and $S$ is the spin quantum number at each site. This expression 
makes clear that, for $J_\times = J_\|$, the Hamiltonian (\ref{eq:essh}) has 
$L$ purely local conservation laws, namely the total rung spin $\vec{T}_i^2$, 
which we may encode in additional quantum numbers $T_i$.

Although the form of Eq.~(\ref{eq:exeh}) is valid for all $S$, we restrict 
our considerations henceforth exclusively to the case $S = 1/2$. Thus $T_i$ 
takes the values $0$ (a rung singlet, indicated by the ellipsoids in 
Fig.~\ref{fig:s}) or $1$ (a rung triplet, represented by the two parallel 
rung spins in Fig.~\ref{fig:s}). For a given configuration $\{T_i\}$, the 
first term of Eq.~(\ref{eq:exeh}) is finite only for groups of $n$ neighboring 
rung triplets ($T_i = 1$) with $n \ge 2$, in which case the Hamiltonian for 
the triplet cluster is that of an open $n$-site spin-$1$ chain. The second 
term in Eq.~(\ref{eq:exeh}) is simply a number operator penalizing the 
presence of these rung triplets relative to rung singlets ($T_i = 0$).

\subsection{Spectrum}
\label{ssec:spec}

\subsubsection{Ground State}
\label{sssec:gs}

As noted first in Ref.~\cite{gelfand91}, the Hamiltonian (\ref{eq:exeh}) of 
the fully frustrated $S = 1/2$ ladder [i.e.~the model of Eq.~(\ref{eq:essh}) 
with $J_\times = J_\|$] possesses a first-order quantum phase transition as a 
function of the coupling ratio $J_\perp / J_\|$ \cite{rx,rhmt}. This transition 
separates a rung-singlet phase for strong $J_\perp$ from a rung-triplet, or 
Haldane \cite{haldane83}, phase at weak $J_\perp$. These two states have been 
found to dominate the phase diagrams of different generalized tetrahedral 
cluster models \cite{Weihong98,Wang00,rbb,MiTo02,Starykh04,rvh,Kim08,HiSta10,
Poilblanc10,lamas15,CCBZ15}. In the formulation of Eq.~(\ref{eq:exeh}), 
the ground states are characterized by all $T_i = 0$ ($n = 0$, rung-singlet 
phase) when $J_\perp$ is dominant or all $T_i = 1$ ($n = L$, Haldane phase) 
when the combination of $J_\|$ and $J_\times$ forces the creation of rung 
triplets to satisfy all of the inter-rung bonds. The ground-state energies 
are simply $E_{n=0} = -\frac{3}{4} \, J_\perp \, L$ for the rung-singlet state 
and $E_{n=L} = {\textstyle \frac{1}{4}} \, J_\perp \, L + E_{S=1}(L)$ for the 
rung-triplet state, where $E_{S=1}(L)$ is the ground-state energy of an 
$L$-site spin-1 chain with coupling constant $J_\|$. By using literature 
estimates \cite{white93,golinelli94} for $e_{\infty} = \lim_{L \to \infty} 
E_{S=1}(L)/L$, the thermodynamic limit for the ground-state energy density 
of the spin-1 chain, one may conclude \cite{rx,rhmt,gelfand91,lamas15} that 
the critical coupling constant for the quantum phase transition is  
\begin{equation} 
J_{\perp,c} = - e_\infty \approx 1.401484\,J_\|  \, .
\label{eq:Jcrit}
\end{equation}
With a view to discussing the thermodynamic properties of the fully 
frustrated ladder, in the remainder of this subsection we analyze the 
spin excitations above the two different ground states.

\subsubsection{Excited States of the Fully Frustrated Ladder and of Open 
Spin-1 Chains}
\label{sssec:es}

The fundamental excitation of the rung-singlet state is a single rung triplet, 
represented schematically in Fig.~\ref{fig:s}. In conventional 1D systems, 
this is a collective mode involving all rungs, and for clarity we refer to it 
henceforth as the triplon. In the fully frustrated ladder, the triplon is a 
non-dispersive mode with a flat band at energy $\omega_k = J_\perp$. Further, 
it has been shown that the $n$-triplet clusters within the rung-singlet state 
form exact bound states \cite{rx}, which are fully localized objects 
\cite{gelfand91}, and therefore are also non-dispersive. Thus one obtains 
an entirely discrete spectrum composed of the levels of $n$-site open Haldane 
chains, separated by an additional constant for the total $n$ of the cluster; 
the case $n = 2$ is discussed in Ref.~\cite{rx}. In the rung-triplet phase, 
the ground state of the Haldane chain is known to be a total singlet with 
complex, system-scale spin entanglement quantified by the string-order 
parameter and dispersive triplet excitations with a gap $\Delta = 0.4105 J_\|$ 
\cite{white93,golinelli94}. These collective modes can be expected in the 
fully frustrated ladder to be accompanied by local excitations due to clusters 
of rung singlets in the triplet background. 

\begin{figure}[t]
\includegraphics[width=\columnwidth]{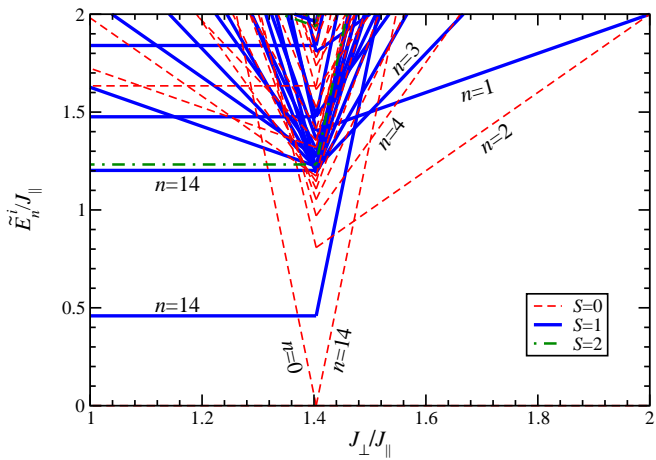}
\caption{(Color online) Low-energy excited levels, ${\tilde E}_n^i$, of a 
14-rung fully frustrated spin-1/2 ladder, shown 
% in units of $J_\times = J_\| = 1$ and 
as a function of $J_\perp/J_\|$. All excitations are classified according 
to their total spin quantum number, $S$, and some by an additional label for 
the number, $n$, of consecutive rungs in the triplet state.}
\label{fig:ED14x2exec}
\end{figure}

We begin a quantitative discussion of the excitations by showing in 
Fig.~\ref{fig:ED14x2exec} the complete low-energy spectrum of a fully 
frustrated 14-rung ladder, with $J_\times = J_\| = 1$ as the unit of energy 
and $J_\perp$ as the variable parameter. This figure was obtained from a 
full diagonalization, which is described in detail in Sec.~\ref{sec:ED}.
Here the rung-triplet phase is on the left side, the rung-singlet one is 
on the right, and the excitation energies, ${\tilde E}_n^i$, are the energy 
differences with respect to the corresponding ground state. We classify all 
the states in Fig.~\ref{fig:ED14x2exec} by their total spin, $S$, and find 
only states with $S \le 2$ in the energy range covered by the figure, of 
which just three, all at high energies, are $S = 2$. We also label some of 
the states by the number $n$ of consecutive triplet rungs ($T_i = 1$) and 
comment that those states containing a single cluster with $n < L$ consecutive 
rung triplets are $L$-fold degenerate, because they may be placed on the 
ladder in $L$ possible ways.

\begin{table}[b]
\begin{tabular}{c|c|c||c|c|c}
 \ $n$\ & spin $S$ & $E^i_{S=1}(n)/J_{\|}$ &
   \ $n$\ & spin $S$ & $E^i_{S=1}(n)/J_{\|}$ \\ \hline
$1$ & $1$ & $0$ 	& $7$  & $1$ & $-8.634532$ \\
$2$ & $0$ & ${- 2}$	&      & $0$ & $-8.303576$ \\
    & $1$ & ${- 1}$	& $8$  & $0$ & $-10.124637$ \\
    & $2$ & ${  1}$	&      & $1$ & $-9.922759$ \\
$3$ & $1$ & ${- 3}$	& $9$  & $1$ & $-11.432932$ \\
    & $0$ & ${- 2}$	&      & $0$ & $-11.220229$ \\
    & $1$ & ${- 1}$	& $10$ & $0$ & $-12.894560$ \\
    & $2$ & ${- 1}$	&      & $1$ & $-12.756229$ \\
    & $1$ & ${  0}$	& $11$ & $1$ & $-14.230359$ \\
    & $2$ & ${  1}$	&      & $0$ & $-14.088587$ \\
    & $3$ & ${  2}$	& $12$ & $0$ & $-15.674010$ \\
$4$ & $0$ & $-4.645751$ &      & $1$ & $-15.576869$ \\
    & $1$ & $-4.136582$ & $13$ & $1$ & $-17.028266$ \\
$5$ & $1$ & $-5.830213$ &      & $0$ & $-16.931557$ \\
    & $0$ & $-5.283567$ & $14$ & $0$ & $-18.459853$ \\
$6$ & $0$ & $-7.370275$ &      & $1$ & $-18.390687$ \\
    & $1$ & $-7.062489$ & $L$  & $0$ & $(-1.401484\ldots)\,L$\\
\end{tabular}
\caption{Energies $E^i_{S=1}(n)$ of open, length-$n$, spin-1 chains with 
exchange constant $J_{\|}$, classified by their total spin, $S$, and labeled 
by their different levels, $i$, in ascending order of energy. The value listed 
for $n = L$ is an extrapolation to the thermodynamic limit, where the boundary 
conditions of the calculation are irrelevant.}
\label{tab:S1energies}
\end{table}

For the interpretation of Fig.~\ref{fig:ED14x2exec}, in Table 
\ref{tab:S1energies} we show the energies of various multiplets found 
in the spectrum of open, $n$-site, spin-1 chains for all values of $n$ up 
to 14. The table is complete up to $n = 3$, beyond which we quote only the 
lowest states, $i = 1$ and 2. Some of this information has been obtained in a number of 
previous studies, specifically the  spectrum for $n \le 2$ \cite{rx},
energy differences between the lowest two levels \cite{kennedy90}, 
and the lowest energy for each $n$ \cite{NUZ97}. Using the Hamiltonian in the form 
of Eq.~(\ref{eq:exeh}), the energy of a single $n$-site, spin-1 cluster 
embedded in a fully frustrated $L$-rung ladder is 
\begin{equation}
E_{n}^i = E^i_{S=1}(n) + (n - {\textstyle \frac{3}{4}} L) \, J_\perp \, ,
\label{eq:ClEn}
\end{equation}
from which it is clear that many (but not all) of the lines in 
Fig.~\ref{fig:ED14x2exec} may be identified by their slope and from the 
data of Table \ref{tab:S1energies}.

\begin{figure}[t]
\centering\includegraphics[width=\columnwidth]{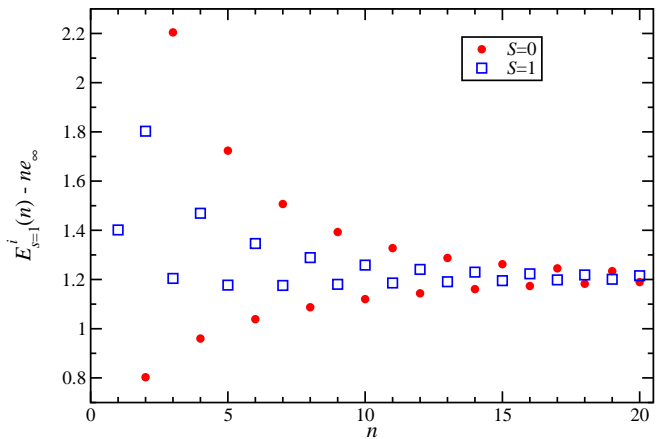}
\caption{(Color online) Energies of the two lowest states of an open, 
$n$-site, spin-1 Heisenberg chain relative to that of an $n$-site segment 
of an infinite chain, shown as a function of $n$ and in units of $J_\|$. Red 
circles denote total singlets ($S = 0$) and open blue squares total triplets 
($S = 1$).}
\label{fig:S1openEn}
\end{figure}

Inspection of Table \ref{tab:S1energies} reveals that the lowest states
of open spin-1 chains have total spins $S = 0$ and $1$; when $n$ is odd, 
the lowest-energy state is $S = 1$, whereas for $n$ even, it has $S = 0$. 
This result may be understood in terms of the valence-bond-solid picture 
\cite{AKLT} of the Haldane phase, where the open chain ends give rise to 
effective $S = 1/2$ spins unable to form valence bonds. These end spins 
experience an effective interaction, mediated by the ``bulk'' of the chain, 
which depends on the Hamiltonian and, for the Heisenberg chain, is such that 
the singlet state of the two end spins is lower in energy for $n$ even, 
i.e.~their effective interaction is antiferromagnetic, whereas for $n$ odd 
the triplet is lower and the interaction ferromagnetic \cite{kennedy90}. To 
display this result in a different perspective, in Fig.~\ref{fig:S1openEn}
we show for $n \le 20$ the energies $E_{S=1}^i(n) - n \, e_{\infty}$ of the two 
lowest states of an open, $n$-site, spin-1 Heisenberg chain relative to the 
energy of the corresponding segment of an infinite chain (cf.\ Ref.~\cite{NUZ97}). For the fully 
frustrated ladder, this quantity gives the excitation energy of a length-$n$ 
triplet segment in a background of rung singlets at the transition point, 
$J_{\perp,c}$ (\ref{eq:Jcrit}), and one observes many of the levels of the 
discrete spectrum making the most important contributions to the physical 
properties of the system.

Figure \ref{fig:S1openEn} also displays the valuable information that the 
cost of breaking one bond in an infinite spin-1 Heisenberg chain is 
\begin{equation}
E_{\rm bond} = \lim_{n \to \infty} (E_{S=1}^i (n) - n \, e_{\infty}) \approx 1.208 
\, J_\| \, ,
\label{eq:S=1bond}
\end{equation}
but that interactions between the two effective end spins lead to significant 
lowering of this energy when the chain segment (triplet cluster) is short.

\subsubsection{Excitations in the Rung-Singlet Phase}
\label{sssec:rs}

The information contained in Figs.~\ref{fig:ED14x2exec} and \ref{fig:S1openEn} 
and in Table \ref{tab:S1energies} may be used to provide a complete discussion 
of the low-energy excitations of the rung-singlet regime, $J_\perp \ge 
J_{\perp,c}$. First we comment that the lowest-lying state visible in 
Fig.~\ref{fig:ED14x2exec} on this side of the transition, the $S = 0$ 
state with $n = 14 = L$, is in fact a finite-size effect. This is the Haldane 
ground state at $J_\perp \le J_{\perp,c}$, which has an energy proportional both 
to the distance from the transition point and to the system size. In the 
thermodynamic limit, its energy for any value $J_\perp$ in the rung-singlet 
phase is infinite and it has no effect on the physical properties.

The energies of all the other states in Fig.~\ref{fig:ED14x2exec} 
depend not on the system size but the cluster size, and these remain 
low-energy excitations in the thermodynamic limit. For $J_\perp > 2 \, 
J_\|$, the lowest excitation is a single rung triplet ($n = 1$), which 
has energy $J_\perp$, can be created on any rung, and creates the 
non-dispersive one-triplon band. At $J_\perp \le 2 \, J_\|$ (the right 
boundary of Fig.~\ref{fig:ED14x2exec}), the $S = 0$ branch of the 
two-triplon bound state ($n = 2$), with a relative energy obtained from 
Eq.~(\ref{eq:ClEn}) of ${\tilde E}_2^1 = E_2^1 + {\textstyle \frac{3}{4}} 
J_\perp L = 2 J_\perp - 2 J_\|$, lies lowest, and remains so until the 
transition, where its gap is approximately $0.803\,J_\|$ (cf.~Ref.~\cite{rx}). 
In this regime of $J_\perp$, the total singlet ($S = 0$) of the ($n = 
4$)-triplon bound state crosses the one-triplon state at $J_\perp \simeq 
1.55 \, J_\|$, as may also be observed in Fig.~\ref{fig:ED14x2exec}. Next, 
the total triplet ($S = 1$) level of the 3-triplon bound state, which is 
the lowest branch of this cluster (Fig.~\ref{fig:S1openEn} and Table 
\ref{tab:S1energies}) crosses at $J_\perp \simeq 1.5\,J_\|$, on its way to 
an energy ${\tilde E}_3^1 \simeq 1.204 \, J_\|$ at the transition. In fact 
the lowest magnetic excitation at $J_\perp = J_{\perp,c}$ is the $S = 1$ branch 
of the $n = 5$ bound state, which has a gap ${\tilde E}_5^1 \simeq 1.18 \, 
J_\|$ there. Thus the triplet gap is almost $50\%$ larger than the singlet 
gap at the transition.

From the standpoint of discussing thermodynamic properties close to the 
quantum phase transition, the most important point in Fig.~\ref{fig:ED14x2exec} 
is the fact that many further states with $S = 0$ and $1$ fall below the 
one-triplon gap as $J_\perp \rightarrow J_{\perp,c}$. These are the lowest-lying 
states of $n$-triplon clusters with all higher values of $n$, and their 
energies exactly at $J_\perp = J_{\perp,c}$ are shown in Fig.~\ref{fig:S1openEn}. 
As $n \to \infty$, these energies converge to the value $E_{\rm bond} \approx 
1.208 \, J_\|$ discussed in Eq.~(\ref{eq:S=1bond}). The rung-singlet phase in 
the regime close to the transition is therefore characterized by a very large 
number of states, of many-triplon origin, whose energies lie close to 
$E_{\rm bond}$, accompanied by some few-triplon bound states providing levels, 
predominantly singlets, at lower energies.

\subsubsection{Excitations in the Rung-Triplet Phase}
\label{sssec:rt}

In the rung-triplet (Haldane) phase at $J_\perp \le J_{\perp,c}$, again the 
ground state of the rung-singlet phase appears in Fig.~\ref{fig:ED14x2exec} as 
a finite-size effect when $J_\perp \rightarrow J_{\perp,c}$. This total singlet, 
labeled with $S = 0$ and $n = 0$, again has a slope proportional to $L$ and 
should be discounted in discussing the infinite system. Horizontal lines
in Fig.~\ref{fig:ED14x2exec} are states with all $n = L = 14$ rungs in a 
spin triplet, and it is clear that not many such states with energies below 
$2\,J_\|$ exist for this system size. The lowest of these lines, located at 
${\tilde E}_{14}^1 = E_{14}^1 - {\textstyle \frac{1}{4}} J_\perp L \simeq 0.46 \, 
J_\|$, is a total-spin triplet ($S = 1$) and corresponds to the ``one-magnon'' 
Haldane gap \cite{haldane83} for $L = 14$; as noted above, the value of this 
gap in the thermodynamic limit is $\Delta \simeq 0.4105 \, J_\|$ 
\cite{white93,golinelli94}. The Haldane chain is also known to have 
two-magnon scattering states with total spins $S = 0$, $1$, and $2$, which 
set in above the threshold energy $2\,\Delta \simeq 0.821\,J_\|$. However, 
the lowest $S = 2$ state for $L = 14$ (Fig.~\ref{fig:ED14x2exec}) 
is found at ${\tilde E}_{14}^2 \simeq 1.23\,J_\|$ and the lowest $S = 0$ 
state at ${\tilde E}_{14}^5 \simeq 1.63\,J_\|$, showing that these extended 
(collective) features of the spectrum in the rung-triplet phase are subject 
to significant finite-size effects.

By contrast, the spectrum for $J_\perp \le J_{\perp,c}$ also contains levels 
with $S = 0$ or $1$ and a non-zero slope in Fig.~\ref{fig:ED14x2exec}. These 
features are the energy branches of rung-singlet clusters ($T_i = 0$) in a
background of rung triplets ($T_i = 1$). Here the rung-singlet number is 
$L - n$, with $n$ rung triplets, and the dominant features are those with 
small $L - n$ ($n$ approaching $L = 14$); as $n$ increases, the 
energies at $J_{\perp,c}$ follow exactly the size-dependence shown in 
Fig.~\ref{fig:S1openEn}, converging for large clusters on the value 
$E_{\rm bond}$ (\ref{eq:S=1bond}). Thus, as on the rung-singlet side, the 
rung-triplet phase in the regime near the transition is dominated by a 
very large number of states appearing around a single, fixed energy close
to (but above) the gap. 

\subsubsection{Summary of the Low-Energy Spectrum}
\label{sssec:ss}

\begin{figure}[t]
\includegraphics[width=\columnwidth]{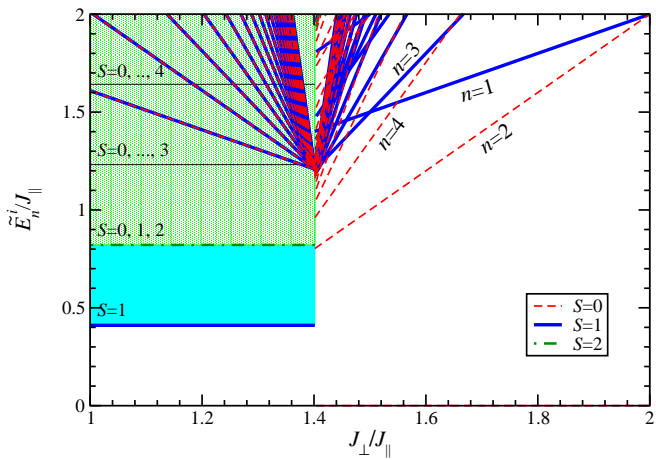}
\caption{(Color online) Schematic overview of the low-energy excitations, 
${\tilde E}_n^i$, of the infinite, fully frustrated spin-1/2 ladder, shown 
%in units of $J_\| = J_\times = 1$ and
as a function of $J_\perp/J_\|$ for a broad 
region around the quantum phase transition at $J_{\perp,c} = 1.410484\,J_\|$. 
Excitations are classified according to their total spin quantum number, $S$. 
The additional label $n$ for the rung-singlet phase specifies the origin of 
excitations in the bound states of $n$-rung triplets. Shaded regions in the 
rung-triplet phase denote single- and multiple-magnon continua.}
\label{fig:LinfAllExec}
\end{figure}

A summary of the previous two subsections for the spectrum of the 
fully frustrated ladder in the thermodynamic limit is provided in 
Fig.~\ref{fig:LinfAllExec}. In the rung-singlet phase ($J_\perp > J_{\perp,c}$), 
the spectrum is purely discrete, with multiple low-energy branches appearing 
well below the one-triplon gap as $J_\perp \rightarrow J_{\perp,c}$. These have 
their origin in some of the shortest clusters, in particular the $S = 0$ 
component of the two-triplon bound state, which provides the lowest excited 
state over the entire region $J_{\perp,c} < J_\perp < 2\,J_\|$.

In the rung-triplet phase ($J_\perp < J_{\perp,c}$), the lowest-lying modes 
are the dispersive single- and multi-magnon bands of the Haldane chain. The 
one-magnon band begins at a threshold $\Delta$ and consists only of $S = 1$ 
excitations. The two-magnon band has a threshold of $2\,\Delta$ and gives 
rise to additional $S = 0$ and $2$ excitations. Two further thresholds fall 
within the energy range of Fig.~\ref{fig:LinfAllExec} and are indicated by
thin horizontal lines; at the three-magnon threshold, $3\,\Delta$, the first 
excitations with total spin $S = 3$ appear, and similarly at $4\,\Delta$. 
The spectrum contains additional discrete (non-dispersive) energy levels due 
to excited rung singlets or rung-singlet clusters; increasingly long clusters 
provide low-lying levels as the transition is approached. As $J_\perp 
\rightarrow J_{\perp,c}$, all of the discrete excitations converge with 
increasing $n$ to a single energy, $E_{\rm bond}$, and are $4L$-fold degenerate 
because the energies of the total $S = 0$ and $1$ states of the two effective 
end spins converge as their interaction vanishes.

To understand the densities of excited states in Fig.~\ref{fig:LinfAllExec}, 
we appeal again to finite ladders. Excitations arising from one cluster of 
rung triplets (singlets) in a background of singlets (triplets) are all 
$L$-fold degenerate for an $L$-rung system. Because there are $L$ such 
states, $L^2$ excitations have energies that converge to $E_{\rm bond} \simeq 
1.208\,J_\|$ at the critical point, and this is identical to the finite-size 
properties of a two-particle continuum. In the rung-triplet phase, the energy
$E_{\rm bond}$ lies just below the three-particle threshold. However, we 
comment that the one-magnon band of the Haldane chain extends up to an energy
of approximately $6 \, \Delta$ \cite{white93}, which is beyond the range of 
Fig.~\ref{fig:LinfAllExec}, and thus the two-magnon continuum, extending to 
$12 \, \Delta$, includes many states lying far above the energies of the 
discrete levels close to $J_{\perp,c}$. Thus we expect the discrete levels to 
provide important contributions to the thermodynamic properties of the fully 
frustrated ladder for all values of $J_{\perp}$ falling within the range of 
Figs.~\ref{fig:ED14x2exec} and \ref{fig:LinfAllExec}, not only on the 
rung-singlet side but also on the rung-triplet side, even if they are 
not the lowest-energy excitations. We analyze these contributions in 
Subsecs.~\ref{ssec:at} and \ref{ssec:anc} by considering short clusters 
and in Subsec.~\ref{ssec:DW} by considering a domain-wall description.

\subsection{Analytical Thermodynamic Approximations}
\label{ssec:at}

To gain some initial insight into the effects of its unconventional spectrum 
on the thermodynamic properties of the fully frustrated ladder, in this 
section we consider a straightforward analytical approximation. Based on  
the exact $n$-triplon bound states of the rung-singlet phase ($J_{\perp} > 
J_{\perp,c}$), we construct the partition functions of $n$-rung clusters and 
use them to obtain the magnetic specific heat, which we denote simply by 
$C(T)$, and magnetic susceptibility, $\chi(T)$, for ladder segments of 
increasing $n$. These results can also be compared with the analogous 
quantities computed for an unfrustrated ladder ($J_\times = 0$) in the regime 
of strong $J_\perp$ to observe the wider implications of strong frustration. 
For the purposes of this analysis, which we apply only in the rung-singlet 
regime, it is convenient in this subsection to normalize the energy scales 
of the system to the rung coupling, $J_\perp$.

The partition function of a single rung, $Z_1$, reflects the four available 
states, namely the singlet and three triplon components. That for a pair of 
rungs, $Z_2$, contains in place of the nine possible states of two separate 
triplons the nine levels of the total $S = 0$, 1, and 2 branches of the 
two-triplon bound state. By continuing in this way for the three-triplon 
bound states of a three-rung cluster, one obtains 
\begin{eqnarray}
Z_1(\beta) & = & 1 + 3 e^{-\beta J}, \nonumber \\
Z_2(\beta) & = & 1 + 6 e^{-\beta J} + \sum_i g_2^i e^{-\beta {\tilde E}_2^i}, 
\label{ez123} \\ 
Z_3(\beta) & = & 1 + 9 e^{-\beta J} + 3 \sum_i g_2^i e^{-\beta {\tilde E}_2^i} 
+ \sum_i g_3^i e^{-\beta {\tilde E}_3^i},  \nonumber
\end{eqnarray}
where $\beta = 1/T$ (we set $k_{\rm B} = 1$) and $g_2^i$ and $g_3^i$ are 
respectively the degeneracies of the different multiplets of the two- 
and three-triplon bound states, whose energies are ${\tilde E}_2^i$ and 
${\tilde E}_3^i$ in the notation of Sec.~\ref{ssec:spec}. Note that the 
combinatorial factor in the expression for $Z_3$ assumes that two-triplon 
bound states are formed for all possible pairs of triplon locations, implying 
that the boundary conditions of the three-rung cluster are periodic rather 
than open, and we will not differentiate between these cases for the 
approximate purposes of the current analysis. In the presence of a magnetic 
field, $h = g \mu_{\rm B} H$, each multiplet is split into all of its separate 
levels, replacing $g_n^i$ in Eq.~(\ref{ez123}) by factors of the form $1 + 2 
\cosh (\beta h) + 2 \cosh (2 \beta h) + ...$ The magnetic specific heat is 
obtained from the free energy, $F = - \beta^{-1} \ln Z$, using the expression 
\begin{equation}
C (T) = - \frac{T}{V} \frac{\partial^2 F(T,h)}{\partial T^2} = 
\frac{\beta^2}{V} \frac{\partial^2 \ln Z (\beta)}{\partial \beta^2} 
\label{eq:cdef}
\end{equation}
and the magnetic susceptibility using
\begin{equation}
\chi (T) = - \lim_{h \rightarrow 0} \frac{1}{V} \frac{\partial^2 F(T,h)}
{\partial h^2} = \lim_{h \rightarrow 0} \frac{1}{\beta V} \frac{\partial^2 
\ln Z(\beta,h)}{\partial h^2}. 
\label{eq:chidef}
\end{equation}
$V$ is the system volume and here we use the number of sites, $V = 2L$, 
quoting all our results per spin-1/2 entity. 

The construction of Eq.~(\ref{ez123}) forms a systematic basis for adding 
the effects of four- and higher-rung triplon clusters to investigate their 
contributions to the thermodynamic properties. %Below
We will also consider 
the thermodynamic quantities derived from $Z_4$, whose multiplet energies, 
${\tilde E}_4^i$, were obtained from the spectrum of the four-site open 
Haldane chain (Table \ref{tab:S1energies}). This method is readily continued 
to larger values of $n$ and its efficacy will be compared with our numerical 
solutions in Sec.~\ref{sec:results}. 

\begin{figure}[t]
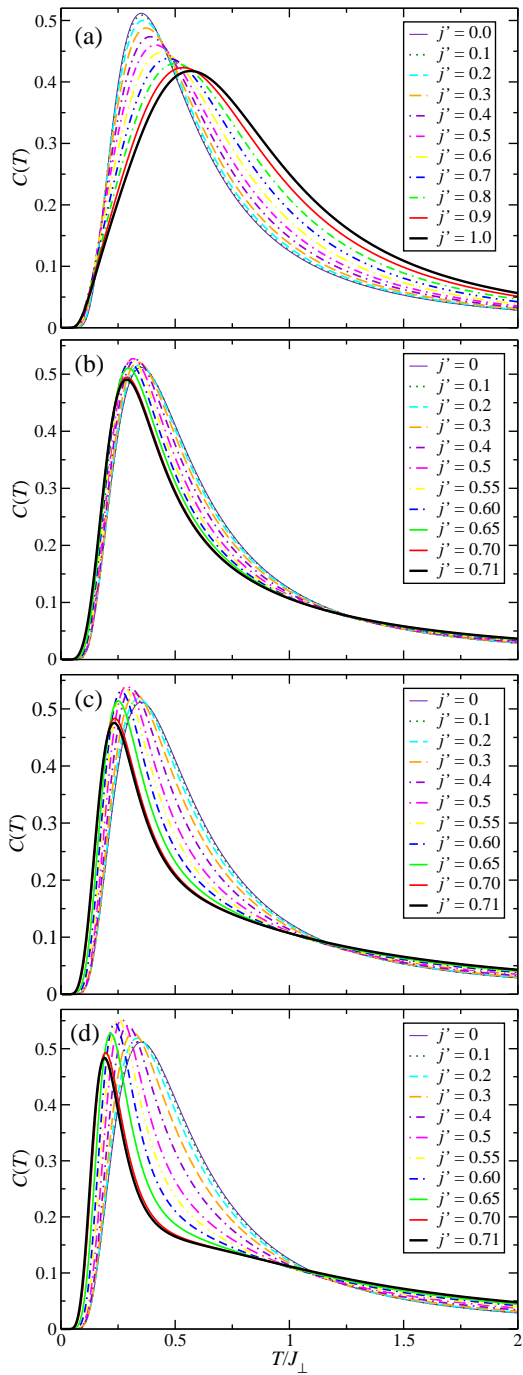

\centering\includegraphics[width=0.38\textwidth]{fulsh}
\centering\includegraphics[width=0.38\textwidth]{fflsh2}
\centering\includegraphics[width=0.38\textwidth]{fflsh3}
\centering\includegraphics[width=0.38\textwidth]{fflsh4t2}
\caption{(Color online) Magnetic specific heat of (a) the unfrustrated 
ladder, (b) two, (c) three, and (d) four fully frustrated rungs, shown 
as functions of temperature for a range of values of the inter-rung 
coupling, $j' = J_\|/J_\perp$. } 
\label{fig:atc}
\end{figure}

Here we focus on the qualitative properties of the specific heat and 
susceptibility, specifically their shape and evolution with the parameters 
of the ladder. We define the quantity $j' = J_\|/J_\perp$ and consider fully 
frustrated ladders ($J_\times = J_\|$) for all values $0 \le j' \le 0.71$ 
within the rung-singlet phase. The thermodynamic properties of the frustrated 
ladder may be benchmarked against those of the unfrustrated two-leg spin 
ladder. These we also calculate by an approximate method, using the 
bond-operator mean-field technique with Ansatz hard-core-boson triplet 
statistics and no higher-order correlations, following Ref.~\cite{rnr}. 
This approach approximates the shifts of spectral weight at finite 
temperatures by a single, effective one-triplon band, which remains sharp 
rather than including thermal broadening. As a result, it is known to 
overestimate the thermal band-narrowing effect, whence the neglect of 
further correlations (which exacerbate this effect). This approximation 
has been found to give an excellent account of the gap and band width of 
the unfrustrated ladder over the parameter range $0 \le j' \le 1$, which 
exceeds the regime of the current investigation. 

Results for the specific heat of unfrustrated ladders and of frustrated 
ladders computed using $Z_2$, $Z_3$ [Eq.~(\ref{ez123})], and $Z_4$ are shown 
in Fig.~\ref{fig:atc} for a range of coupling ratios, $j' = J_\|/J_\perp$, in 
the rung-singlet phase. The common feature of every panel is the curve for 
$j' = 0$, the specific heat (per spin) of an isolated ladder rung, 
\begin{equation}
C_{\rm r} (T) = \frac{3 \, (\beta J_\perp)^2 {\rm e}^{-\beta J_\perp}}{2 \, 
\left(1 + 3 \, {\rm e}^{-\beta J_\perp} \right)^2} \, ,
\label{eq:Cdimer}
\end{equation}
which shows an exponential increase at low temperature characteristic of a 
gap $\Delta = J_\perp$, a maximum at $T = 0.35 J_\perp$, and a power-law decay 
towards higher temperatures. The results for the unfrustrated ladder in 
Fig.~\ref{fig:atc}(a) show that, as the leg coupling $J_\|$ is increased, 
the increasing triplet band width causes a slow but continuous reduction of 
the gap and a shift of weight to higher energies. The specific-heat maximum 
displays a monotonic decrease in its height, $C_{\rm max}$, but an increase 
in the temperature, $T^C_{\rm max}$, at which it occurs, as $j'$ increases. We 
stress that this calculation shows only the contributions of the one-triplon 
excitations, whose density of states is concentrated between a gap $\Delta$, 
which drops from $J_\perp$ at $j' = 0$ to approximately $J_\perp/2$ at $j' = 1$ 
\cite{Frischmuth96,Greven96}, and an upper edge close to $J_\perp + 2 J_\|$. 
However, because the maximum in $C(T)$ already increases with $j'$ for the 
single-band contribution, we expect that multi-triplon contributions, which 
appear at higher temperatures, can only strengthen this effect.

In the fully frustrated ladder, the evolution of the specific heat is 
quite different. For the two-rung approximation [Fig.~\ref{fig:atc}(b)], 
$T^C_{\rm max}$ does not rise with the coupling but instead is suppressed 
monotonically in temperature by increasing $j'$. This is not a consequence 
of changes in the gap, as $\Delta$ begins to drop only beyond $j' = 0.5$ 
(Sec.~\ref{ssec:spec}). $C_{\rm max}$ first increases weakly with $j'$, to 
a maximum value at $j' = 0.5$, before falling somewhat more rapidly as 
$j' \rightarrow j'_c$. The three-rung approximation, shown in 
Fig.~\ref{fig:atc}(c), exaggerates these tendencies, and the four-rung 
approximation exaggerates them still further [Fig.~\ref{fig:atc}(d)], 
meaning that the strongest alterations visible due to including larger 
$n$-triplon clusters occur for the coupling values closest to the 
transition. 

We draw three conclusions from the specific heat. First, its behavior 
as a function of $j'$ in the fully frustrated ladder is completely different 
from that of the unfrustrated ladder over the whole range of $j'$. 
Second, even the two-rung approximation may already capture the basic 
phenomenology of bound-state effects, although additional bound states are 
responsible for strengthening these effects, which are most pronounced close 
to the quantum phase transition at $j'_c$. Third, the peak of the specific 
heat becomes significantly smaller (narrower as well as lower) as $j 
\rightarrow j'_c$, with very little weight moving to lower energies due to 
the (weakly) decreasing gap; clearly most of the ``missing'' entropy caused 
by the peak suppression near $j'_c$ is retrieved at higher temperatures, 
beyond the apparent crossing point of the curves at $T \approx J_\perp$. 
Thus despite the fact that the one-triplon band is completely flat and the 
leading effect of bound-state formation to produce a sharper peak at lower 
energies, one overall effect of the frustrated coupling is to push the 
available states to a higher average energy. 

\begin{figure}[t]
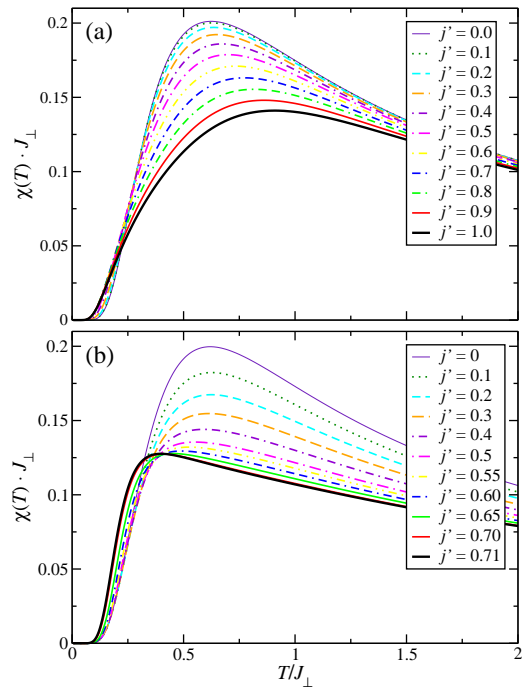

\centering\includegraphics[width=0.38\textwidth]{fulss}
\centering\includegraphics[width=0.38\textwidth]{fflss4t2}
\caption{(Color online) Magnetic susceptibility of (a) the unfrustrated 
ladder and (b) the four-rung approximation to the fully frustrated ladder, 
shown as functions of temperature for a range of values of the inter-rung 
coupling, $j' = J_\|/J_\perp$.}
\label{fig:atchi}
\end{figure}

The corresponding susceptibilities per spin are presented in 
Fig.~\ref{fig:atchi}, where we show only the four-rung approximation 
for the fully frustrated ladder. Again the two panels are anchored by the 
$j' = 0$ result, the susceptibility, 
\begin{equation}
\chi_{\rm r} (T) = \frac{\beta {\rm e}^{-\beta J_\perp}}{\left(1 + 3 \, {\rm 
e}^{-\beta J_\perp} \right)} \, ,
\label{eq:ChiDimer}
\end{equation}
of an isolated spin-1/2 dimer with coupling constant $J_\perp$ 
\cite{BB52,Johnston00a,Deisenhofer06}, which is activated at low $T$ by the 
gap $\Delta = J_\perp$, has a broad peak at approximately $T = 0.62 J_\perp$, 
and falls very slowly to high temperatures. The susceptibility of the 
unfrustrated ladder [Fig.~\ref{fig:atchi}(a)] has the same basic features 
as the specific heat [Fig.~\ref{fig:atc}(a)], namely that the gap determining 
the low-$T$ response decreases steadily with $j'$ but the peak position,
$T^{\chi}_{\rm max}$, increases, although its height, $\chi_{\rm max}$, falls.  
The two quantities $C$ and $\chi$ are always characterized by the same gap. 

The susceptibility of the fully frustrated ladder [Fig.~\ref{fig:atchi}(b)] 
is again quite different from the unfrustrated case, but not particularly 
similar to the specific heat [Fig.~\ref{fig:atc}(d)]. $T^{\chi}_{\rm max}$ 
barely moves with $j'$ until $j' > 0.3$, then moves increasingly rapidly 
to lower temperatures; $\chi_{\rm max}$ does fall with $j'$ but then changes 
very little as $j'$ approaches $j'_c$. We remind the reader that, while $C$ 
is the response due to all excited states, $\chi$ is the response due only 
to all excited magnetic states, and thus the gap to triplet and higher-spin 
excitations is constant ($\Delta^{\chi} = J_\perp$) for all values of $j'$ 
except for a small drop over the range $2/3 \le j' \le j'_c$, ending at 
$\Delta_c^{\chi} = 0.857 J_\perp$. Thus we have the situation that $\chi$ 
is not characterized by the same gap as $C$, which is quite unusual in 
magnetic systems and is another consequence of the strong frustration. 

The susceptibilities calculated using $Z_2$ and $Z_3$ (not shown here, but 
presented for selected $j'$ values in Sec.~\ref{ssec:anc}) display 
the same trend towards the features of the $Z_4$ result as do the cruder 
approximations to the specific heat (Fig.~\ref{fig:atc}). Again one may 
conclude that very small numbers of bound states are sufficient to capture 
all the features of the thermodynamic response when $j'$ is far from $j'_c$ 
in the rung-singlet phase. We will quantify this statement by comparison 
with our numerical results in Sec.~\ref{sec:interp}, where we will 
find that the four-rung approximation of Figs.~\ref{fig:atc}(d) and 
\ref{fig:atchi}(b) already achieves quantitative accuracy for $j' \le 0.5$, 
i.e.~contributions from $n$-triplon clusters with $n > 4$ are negligible 
in this regime. The crucial effects of bound-state formation in the fully 
frustrated ladder that we wish to investigate with our advanced numerical 
techniques are those occurring in the vicinity of the quantum phase 
transition, and thus in Secs.~\ref{sec:methods} and \ref{sec:results} 
we will focus on the parameter range $0.5 \le j' \le 1$ shown in 
Figs.~\ref{fig:ED14x2exec} and \ref{fig:LinfAllExec}.

\section{Numerical Methods}
\label{sec:methods}

The Hamiltonian of the fully frustrated ladder expressed in the rung basis 
of Eq.~(\ref{eq:exeh}) is particularly valuable from a numerical standpoint.
In this section we explain the methods we apply both to extend ED calculations 
to larger system sizes and to perform sign-problem-free QMC simulations even 
for a highly frustrated system. Readers not interested in the technical 
details of these approaches may continue directly to the results presented 
in Sec.~\ref{sec:results}.

\subsection{Exact Diagonalization}
\label{sec:ED}

The model specified in the form (\ref{eq:exeh}) has a large number of sectors
classified by the fixed set of rung quantum numbers $\{T_i\}$, $i = 1$, 
$\ldots$, $L$, and each sector may be diagonalized separately. Further, as 
explained in Sec.~\ref{ssec:spec}, every sector can be characterized entirely 
by either the spectrum of a periodic, $L$-site, spin-1 chain (if all rungs are 
in a triplet state) or the spectrum of open, $n$-site, spin-1 chains for all 
$n$-triplon clusters with $n < L$. The complete spectrum is constructed by 
considering the combinatorial factors obtained by embedding all such open 
chain segments into the ladder and counting the (discrete) energy levels 
that appear with their corresponding multiplicities. 

For the two cases where $T_i = 0$ for all $i$ (rung-singlet phase) or $T_i = 1$ 
for all $i$ (rung-triplet), there is only one such possibility. Single chain 
segments, of any $n$, may be embedded at $L$ places and therefore the energies 
(\ref{eq:ClEn}) are $L$-fold degenerate (Sec.~\ref{sec:ffl}). For general 
numbers of chain segments, of different lengths and positions on the ladder,
the combinatorial factors are enumerated by computer. The key physical 
property allowing such a straightforward approach is that the interaction 
between any two neighboring singlet and triplet rungs (and indeed between 
two singlet rungs) is precisely zero, so that rung and cluster states need 
only be placed side by side in all possible combinations. This type of 
strategy has been applied previously \cite{DRHS07,HHPR11,ohanyan12} to
models with local conservation laws \cite{TKS96,NUZ97,NUZ98,RIS98,KON00,SchuRi02},
including to the fully frustrated ladder \cite{DRHS07},
where the calculations were restricted to systems up to 
$L = 12$ (24 spins). Here we extend this treatment to $L = 14$, a value 
that we note lies beyond the limit of 24 $S = 1/2$ spins accessible in 
conventional full-diagonalization calculations, even for systems with
high spatial symmetry \cite{HMHV06}.

In the calculations presented here, we exploit spatial reflection symmetry 
and also translational symmetry for the periodic $L$-site $S = 1$ system. 
We make use of the conservation of total $S^z$, of spin inversion symmetry, 
and of SU(2) symmetry in order to reconstruct the $S^z = 1$ sector from 
the other sectors. The most demanding computation is the diagonalization of 
the open spin-1 chain with $n = L - 1$ sites. For the $L = 14$ ladder, the 
open spin-1 chain with $n = 13$ sites (representing the contribution of 
13-triplon clusters) has a vector-space dimension up to $85616$ for $S^z = 2$.
Results for the periodic $L = 14$ spin-1 chain are in fact available from a 
previous investigation \cite{cav2o4}. Beyond these full-diagonalization 
procedures, we have also computed the ground-state energies of open spin-1 
chains with $14 \le n \le 20$ sites using the Lanczos algorithm 
\cite{Lanczos,DagottoRevModPhys66}; these results are included in 
Fig.~\ref{fig:S1openEn}, and also in Table \ref{tab:S1energies} for $n = 14$. 

\subsection{Quantum Monte Carlo}
\label{sec:QMC}

\begin{figure}[t]
\centering
\includegraphics[width=0.9\columnwidth]{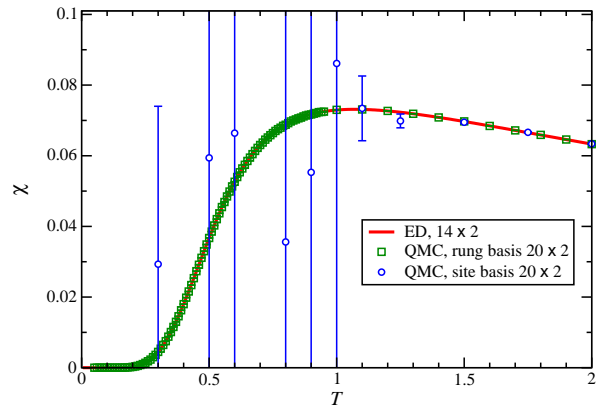}
\caption{(Color online) Comparison of the magnetic susceptibility, $\chi$, 
obtained for a fully frustrated ladder with $J_\perp = 2$ and $J_\| = J_\times
 = 1$, between QMC results for a system of $L = 20$ rungs and ED results for 
$L = 14$. QMC simulations were performed in both the single-site basis and 
the rung basis.}
\label{fig:chiJR2QMCtest}
\end{figure}

Calculations for the thermodynamic properties of all systems larger than 
those accessible by ED require a different numerical method. In 1D the 
primary options for general models are finite-temperature variants of
the DMRG method (Sec.~\ref{sec:intro}) and QMC. Here we employ the latter.
%also as a means to explore how far it can be pushed in situations such
%as the present one.

QMC simulations of frustrated spin systems are well 
known to suffer from the notorious ``sign problem,'' where the statistical 
weights of the effective classical system may become negative. As an example 
we consider a conventional simulation of the fully frustrated ladder in the 
single-site basis of Eq.~(\ref{eq:essh}) using the stochastic series expansion 
(SSE) representation with directed loop updates \cite{Sandvik02}. Results for 
the susceptibility, $\chi$, of an $L = 20$ ladder with $J_\perp = 2\,J_\|$ 
are shown by the open (blue) circles in Fig.~\ref{fig:chiJR2QMCtest}, 
where they are compared with ED results obtained for $L = 14$, a system size
large enough to be considered representative of the thermodynamic limit
for these parameters (as we show explicitly in Sec.~\ref{sec:results}).
Clearly the QMC results obtained in the single-site basis are accurate only 
at high temperatures, but break down even on approaching the maximum of 
$\chi$ at $T \approx J_\|$; specifically, the error bars become so large 
that the maximum cannot be resolved. 

\begin{figure}[t]
\includegraphics[width=0.9\columnwidth]{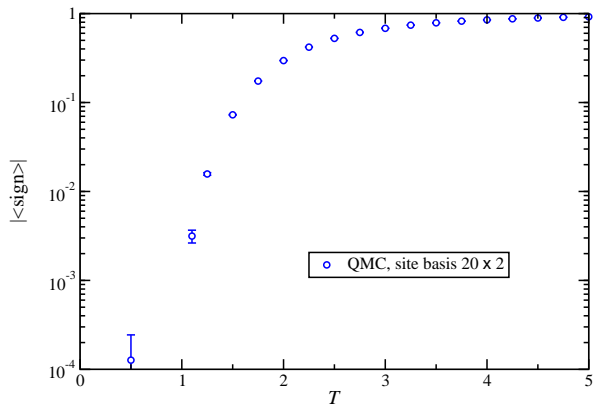}
\caption{(Color online) Average sign obtained during QMC simulations in the 
single-site basis of the $L = 20$ ladder with $J_\perp = 2$ and $J_\times = J_\|
 = 1$.}
\label{fig:signJR2QMCtest}
\end{figure}

This manifestation of the sign problem (Fig.~\ref{fig:chiJR2QMCtest}) is 
presented in Fig.~\ref{fig:signJR2QMCtest} in the form of the average sign, 
$\langle {\rm sign} \rangle$, of the statistical weight obtained at different 
temperatures. At high $T$ this is close to $1$, whereas for temperatures 
$T < 2 J_\|$, and particularly $T < J_\|$, it falls rapidly towards zero (note 
the logarithmic scale in Fig.~\ref{fig:signJR2QMCtest}). Only for $T \gtrsim 
1.5 \, J_\|$ can accurate data be obtained, at a high cost in CPU time, but 
for temperatures $T \lesssim J_\|$ the positive and negative contributions 
are so close in absolute value that the error in their difference can no 
longer be controlled. We comment that $L = 20$ (Figs.~\ref{fig:chiJR2QMCtest} 
and \ref{fig:signJR2QMCtest}) remains a comparatively small system, while the 
sign problem is known to scale exponentially in both system size and inverse 
temperature ($\beta = 1/T$) \cite{TW05}.

However, the nature and severity of the sign problem depend on the basis in 
which the simulation is constructed. The fully frustrated ladder presents a 
particularly clear example where the Hamiltonian can be expressed not in the 
site basis of Eq.~(\ref{eq:essh}) but in the spin-dimer (rung) basis of 
Eq.~(\ref{eq:exeh}). For a given rung $i$, we consider the local basis states
\begin{eqnarray}
|S \rangle_{i} & = & {\textstyle \frac{1}{\sqrt{2}}} ( |\uparrow \downarrow 
\rangle_{i} - |\downarrow \uparrow \rangle_{i}), \nonumber \\
|0 \rangle_{i} & = & {\textstyle \frac{1}{\sqrt{2}}} ( |\uparrow \downarrow 
\rangle_{i} + |\downarrow \uparrow \rangle_{i}), \nonumber \\
|+ \rangle_{i} & = & |\uparrow \uparrow \rangle_{i}, \nonumber \\
|- \rangle_{i} & = & |\downarrow \downarrow \rangle_{i}.
\label{eq:sdb}
\end{eqnarray}
The idea of performing QMC simulations in a spin-dimer basis was suggested by 
Nakamura~\cite{nakamura98}, who considered a different basis, symmetric under 
spin reflection, which allowed him to simulate the $J_1$-$J_2$ Heisenberg 
spin chain over an extended parameter region. Here, however, the basis of 
Eq.~(\ref{eq:sdb}) is more suitable because it makes the total dimer spin 
operators, $T^z_i$, diagonal and thus quantities such as the susceptibility 
are readily accessible.

The change of computational basis requires the simulation of composite-spin 
operators, i.e.~updates of both quantum numbers $T_i$ and $T_i^z$. The 
off-diagonal part of the Hamiltonian in the rung basis (\ref{eq:exeh}) 
can be decoupled in terms of the total-spin raising and lowering operators, 
$T^{\pm}_i$, and the SSE QMC method implemented with the standard 
diagonal update scheme \cite{Sandvik02}. Directed loop updates are implemented 
by employing the linear programming approach of Ref.~\cite{alps-sse} to solve 
the directed loop equations numerically in order to minimize bounces during the 
operator loop construction. We considered in particular the set of non-trivial 
permutations of the four spin-dimer basis states as local operators that are 
applied during the directed loop update. For this one may employ both the 
total-spin raising and lowering operators, $T^{\pm}_i$, and the raising and 
lowering operators $D^{\pm}_i$, constructed from the spin-difference operator 
$\vec{D}_i = \vec{S}_i^1 - \vec{S}_i^2$, combined with the local $D^z_i$ 
operators. The latter are not diagonal in the spin-dimer basis. The action 
of the different operators on the local dimer basis states is shown in 
Table \ref{tab:matele}.

\begin{table}[b!]
\begin{tabular}{ | c || c | c | c | c | c | c | c |}
\hline
& $\vec{T}^2_{i}$ & $T^z_{i}$ & $T^+_{i}$       &  $T^-_{i}$        &  $D^z_{i}$  
& $D^+_{i}$        & $D^-_{i}$ \\
\hline
$| S \rangle_{i}$ & 0     & 0  & 0   &  0  &  $| 0 \rangle_{i}$   & $-\sqrt{2}|
 + \rangle_{i}$            & $\sqrt{2}| - \rangle_{i}$ \\
$| 0 \rangle_{i}$ & 2           & 0     & $\sqrt{2}| + \rangle_{i}$          
&  $\sqrt{2}| - \rangle_{i}$    &  $| S \rangle_{i}$   & 0       & 0 \\
$| + \rangle_{i}$ & 2     & 1     & 0           &  $\sqrt{2}| 0 \rangle_{i}$ 
&  0        & 0            & $-\sqrt{2}| S \rangle_{i}$ \\
$| - \rangle_{i}$ & 2           & -1    & $\sqrt{2}| 0 \rangle_{i}$    &  0  
&  0          & $\sqrt{2}| S \rangle_{i}$            & 0 \\
\hline
\end{tabular}
\caption{Action of local total-spin and spin-difference operators on the local 
spin-dimer basis states. Because $\vec{T}^2_{i}$ and $T^z_{i}$ are diagonal  
in this basis, we give only the eigenvalues for these operators.}
\label{tab:matele}
\end{table}

When QMC simulations are performed in the spin-dimer basis, the sign problem 
is completely absent for an even number $L$ of rungs. One may verify that the 
average sign computed in Fig.~\ref{fig:signJR2QMCtest} is now identically equal 
to 1. This occurs because the chain of rung ``super-sites'' forms a bipartite 
lattice on which the exchange interactions between nearest-neighbor total 
spins ($\vec{T}_i$) are bilinear (\ref{eq:exeh}). By using a similar amount 
of CPU time for simulations in the dimer basis as for the single-site 
basis, we now obtain the data shown by the open (green) squares in 
Fig.~\ref{fig:chiJR2QMCtest}. The results resolve not only the maximum of 
$\chi$ but also the complete low-temperature behavior all the way to $T = 0$. 
However, here we find a different consequence of the extensive number of local 
conservation laws in Eq.~(\ref{eq:exeh}), namely that these hinder an effective 
mixing of the local spin states. As a result, the autocorrelation and 
thermalization times become very large at low temperatures and close to 
the quantum phase transition [Eq.~(\ref{eq:Jcrit})], requiring further 
modification of our approach. 

To perform QMC simulations for suitably large system sizes under these 
circumstances, we employed a standard parallel tempering protocol 
\cite{pt1,pt2,pt3} over a set of 64 temperatures, although 32 values were 
found to be sufficient at coupling ratios far from the phase transition. Each 
replica was passed through an initial annealing phase, where it was heated 
slowly from an appropriate low-temperature state to its target temperature. The 
simulations were initialized by taking as the SSE base state (with an empty 
operator string) a rung-singlet product state for $J_{\perp}$ within the 
rung-singlet phase ($J_{\perp} > J_{\perp,c}$) and an antiferromagnetic N\'eel 
state for $J_{\perp}$ within the rung-triplet phase ($J_{\perp} < J_{\perp,c}$). 
The parallel tempering swap rates were adapted to allow for sufficient 
autocorrelation of the replicas between swaps.  

We comment here that a dimer-basis description appears natural for quantum 
magnets with a dimerization in their interaction geometry. The QMC simulation 
scheme presented here can be extended to the case of arbitrary spin systems 
whose interactions are bilinear when reexpressed in the basis of spin-dimer 
singlet and triplet states. In addition to interdimer exchange couplings based 
on $\vec{T}_{i}$ operators, which we denote as $TT$-type terms, the most 
general case would also include couplings of $DD$, $TD$, and $DT$ types. For 
models with $TT$ terms only, and for which the dimer super-sites form a 
bipartite lattice, the QMC sign problem is completely eliminated, as in the 
fully frustrated spin ladder. In fact all such models are strongly constrained 
by local conservation laws on every spin dimer, but any additional interaction 
terms break these laws explicitly. Local transitions then become possible 
between the singlet and triplet sectors, allowing for a more efficient mixing 
of states within QMC simulations, but at the price that such additional 
interaction terms lead generically to a QMC sign problem (even in the 
spin-dimer basis). 

In Table~\ref{tab:matele}, we find that (i) the $\vec{T}_{i}$ operators do 
not create transitions between singlet and triplet sectors and (ii) the 
$D^z_i$ operators do swap local $|S \rangle$ and $|0 \rangle$ states. In 
a model containing only $TT$ and $D^zD^z$ terms, and with a bipartite 
structure of couplings between dimers on different sublattices only, 
again the sign problem is completely absent, because the periodic boundary 
conditions in the discrete imaginary-time propagation direction mandate 
that only even numbers of interdimer $D^zD^z$ terms can create allowed QMC 
configurations. All allowed configurations then contribute to the quantum 
partition function with positive weight (this is a straightforward 
generalization of the reason that no sign problem plagues world-line 
simulations of the antiferromagnetic Heisenberg model on a bipartite 
lattice) \cite{radp}. However, we note that in such a model SU(2) symmetry 
is broken explicitly.

These considerations highlight that special cases exist for particular 
subsets of interaction terms, which can be handled by QMC without generating 
a sign problem at all. For more general systems with dimerized ground states 
\cite{DRHS07,HHPR11,ohanyan12,TKS96,NUZ97,NUZ98,RIS98,KON00,SchuRi02},
the possibility arises that, even when 
non-positive QMC weights can occur, a simulation may have only a mild QMC 
sign problem because the ground state is a direct product state (of dimer 
singlets). Further analysis to assess the full potential of the 
spin-dimer-based approaches to QMC lies beyond the scope of this 
study. We highlight only one further extension of these ideas, to consider 
the construction of more general computational bases, formed from larger 
lattice units, or simplices, such as triangles \cite{rhmt} or four-site 
plaquettes \cite{Mambrini99}. The development of simplex-QMC methods to 
explore geometries and interactions providing sign-problem-free or 
sign-problem-suppressed simulations can be expected to provide valuable 
progress in frustrated quantum magnetism.

\section{Thermodynamic Properties: Numerical Results}
\label{sec:results}

In this section we present our numerical results for the thermodynamic 
properties of the fully frustrated ladder and of some unfrustrated comparison 
cases. We begin by ensuring the accuracy and reliability of our calculations 
by comparing the results from ED and QMC, primarily to verify that finite-size
effects are under control throughout the phase diagram, including in the 
vicinity of the critical point [Eq.~(\ref{eq:Jcrit})]. We present the basic 
phenomenology of the magnetic specific heat, $C$, and susceptibility, $\chi$,  
in terms of the physical features emerging as the interaction parameters are 
changed in two directions, from rung-singlet to rung-triplet ladders and 
from fully frustrated to unfrustrated ladders. Following the results of 
Sec.~\ref{ssec:at}, we focus on the parameter regime $1 \le J_\perp/J_\| \le 2$ 
around the phase transition. We defer to Sec.~\ref{sec:interp} a deeper 
analysis of our results and of their interpretation using the information 
about the spectrum presented in Sec.~\ref{ssec:spec}. However, for the most 
elementary understanding of the response functions it is worth recalling that 
the specific heat is a consequence of all states and therefore reflects 
primarily the singlet bound states in the vicinity of $J_{\perp,c}$, whereas 
the susceptibility is a consequence of all magnetic states and therefore 
reflects primarily triplet bound states. 

\subsection{Finite-Size Convergence: Comparison of ED and QMC}
\label{sec:FinSize}

\begin{figure*}[p]
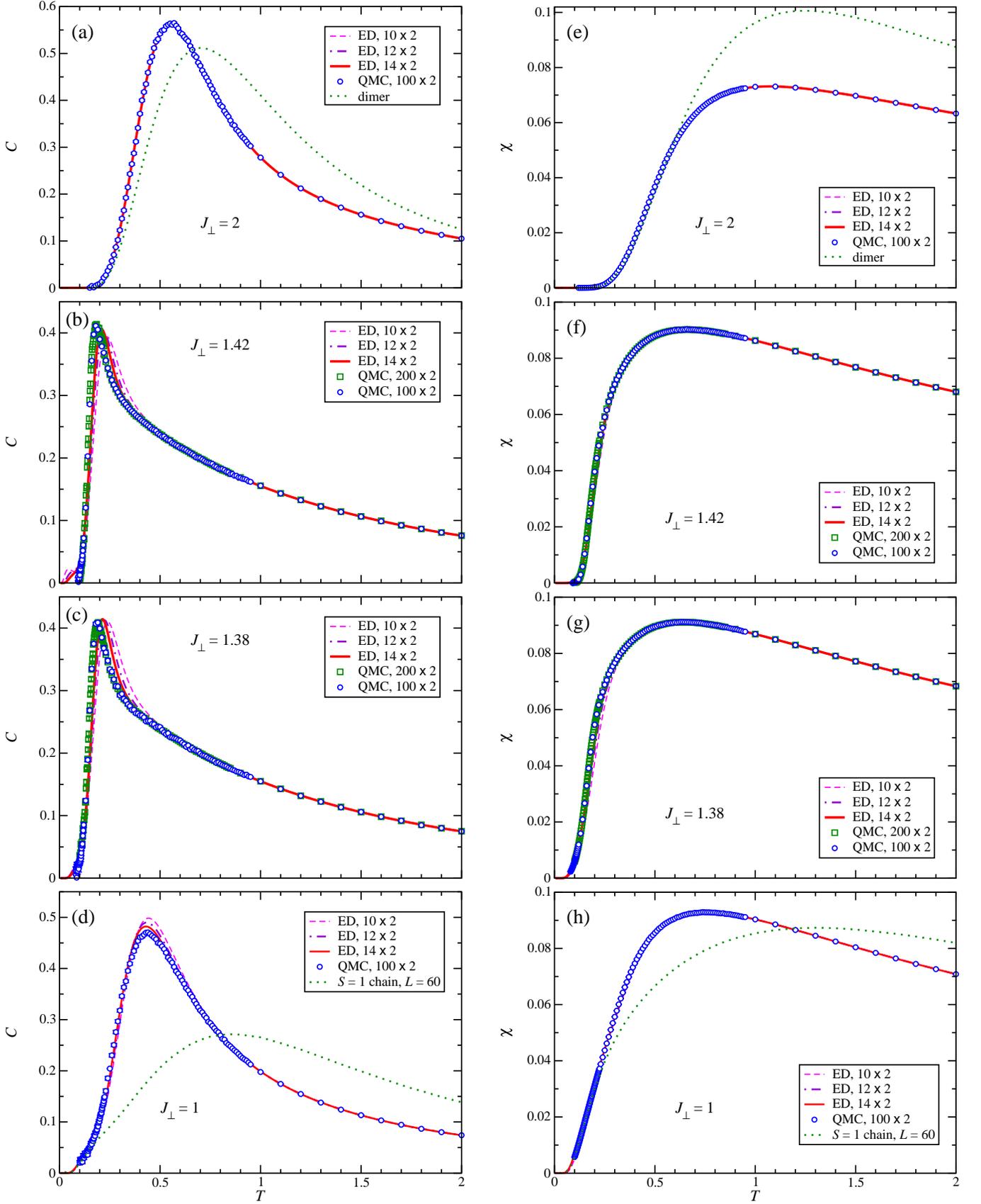

\includegraphics[width=\columnwidth]{CJR2}\hfill%
\includegraphics[width=\columnwidth]{chiJR2}\\
\includegraphics[width=\columnwidth]{CJR1_42}\hfill%
\includegraphics[width=\columnwidth]{chiJR1_42}\\
\includegraphics[width=\columnwidth]{CJR1_38}\hfill%
\includegraphics[width=\columnwidth]{chiJR1_38}\\
\includegraphics[width=\columnwidth]{CJR1}\hfill%
\includegraphics[width=\columnwidth]{chiJR1}
\caption{(Color online) Specific heat, $C$ (left column), and susceptibility, 
$\chi$ (right column), per spin for fully frustrated ladders with inter-rung 
couplings $J_\times = J_\| = 1$ and rung couplings $J_\perp = 2$, $1.42$, $1.38$,
and $1$ (from top to bottom). Each panel compares ED results for system sizes 
of $L = 10$, $12$, and $14$ rungs with QMC results obtained for $L = 100$ 
rungs. In panels (a) and (e) we include the results (\ref{eq:Cdimer}) and 
(\ref{eq:ChiDimer}) for isolated dimers of rung coupling $J_\perp = 2 \, J_\|$. 
In panels (d) and (h) we include QMC results for a spin-1 chain of $L = 60$ 
sites \cite{cav2o4}, normalized to twice the number of spins in the chain.}
\label{fig:CompEDQMC}
\end{figure*}

With a view to benchmarking the quality of our numerical calculations, 
Fig.~\ref{fig:CompEDQMC} compares ED and QMC results for the specific heat and 
susceptibility of fully frustrated ladders with four different values of the 
ratio $J_\perp/J_\|$, selected in the rung-singlet phase far from $J_{\perp,c}$, 
in the rung-singlet phase close to $J_{\perp,c}$, and similarly in the 
rung-triplet phase. Every panel shows ED results calculated for ladders 
of 10, 12, and 14 rungs, along with QMC data for 100-rung ladders and 
additional QMC results from 200-rung ladders when $J_{\perp}$ is close to 
$J_{\perp,c}$. Quite generally, all the susceptibility curves are nearly 
indistinguishable, meaning that finite-size effects in $\chi$ are extremely 
small, and we focus most of our comments on the specific-heat results. 

The fact that identical curves are obtained for the specific heat of the 
$J_\perp/J_\| = 2$ ladder (top) indicates that a system with $L = 10$ rungs 
can be considered as a good approximation to the thermodynamic limit. By 
contrast, finite-size effects are significantly more important at $J_\perp/J_\|
 = 1.42$, where there is a systematic shift of the peak to the left from 
$L = 10$ to 12 to 14, and none of the ED calculations capture the peak position 
of the QMC results. Although this type of shift could be interpreted as 
reflecting the importance of longer clusters, in the same way as observed in 
Fig.~\ref{fig:atc}, it may seem surprising that clusters of more than 10 rungs 
could make a discernible difference at $\left| J_\perp - J_{\perp,c} \right|/J_\| 
\approx 0.02$, given that the ground state is still a simple product of rung 
singlets. However, we recall that in systems of the sizes accessible by ED, 
the lowest excitation in this regime is actually the ``intruder'' state, 
which is the ground state on the rung-triplet side of the transition, as shown 
in Fig.~\ref{fig:ED14x2exec} and discussed in Sec.~\ref{ssec:spec}, and this 
causes differences in the thermodynamic response. Here we also include QMC 
results for a ladder of $L = 200$ rungs, and these make clear that all 
finite-size phenomena are thoroughly suppressed in our $L = 100$ data. 

Turning to the rung-triplet side of the transition, the same finite-size 
effects are visible at $J_\perp/J_\| = 1.38$ as at 1.42; in fact they are 
slightly more pronounced, to the extent that they are visible in the 
corresponding curve for the susceptibility, where we draw attention to 
the different temperatures at the center of the rapid rise. Thus it is 
clear that the thermodynamic response of the Haldane phase close to the 
critical point is significantly affected by the intruding rung-singlet 
ground state. Whether large clusters (of rung singlets) or the one-magnon 
band of the Haldane chain play a role here is unclear. Moving well into the 
Haldane phase, at $J_\perp/J_\| = 1$ we observe finite-size effects not in 
the position of the specific-heat peak at $T \approx 0.5\,J_\|$ but in its 
height. This type of size effect can be found in calculations of the Haldane 
(spin-1 Heisenberg) chain \cite{cav2o4}, and thus its appearance in the 
rung-triplet phase of the ladder suggests that it is due to long spin-1 chain 
segments separated by excited rung singlets. From our results and also from 
those for the Haldane chain, the QMC data for $L = 100$ can be considered as 
fully converged to the thermodynamic limit for this choice of parameters. 
For thermodynamic purposes, we take finite-size effects to be negligible 
in the fully frustrated ladder of $L = 100$ rungs for all values of $J_\perp$.

\subsection{Dependence of $C$ and $\chi$ on $J_\perp$}
\label{ssec:jpd}

We turn now to a discussion of how the thermodynamic response of the 
fully frustrated ladder depends on the rung coupling ratio $J_\perp/J_\|$, 
continuing with accurate numerical data from near the transition the 
investigation begun in the rung-singlet phase in Sec.~\ref{ssec:at}. 
Considering again the sequence of panels in Fig.~\ref{fig:CompEDQMC}, 
at the top is the case $J_\perp = 2\,J_\|$, representative of strong rung 
coupling. As in Sec.~\ref{ssec:at}, we compare the specific heat and 
susceptibility with the results (\ref{eq:Cdimer}) and (\ref{eq:ChiDimer}) 
for decoupled dimers ($J_\perp \gg J_\|$), illustrated here for a bond 
strength $J_\perp = 2$. We observe again that, despite the formation of 
bound states at and above the one-triplon energy (Sec.~\ref{ssec:spec}), 
the leading effect of frustration is to push the maximum of the specific 
heat to a lower temperature and to make its peak higher and narrower than 
that of one dimer. The maximum of the susceptibility is also pushed to lower 
temperatures, although here the dominant effect is the suppression of the 
peak height compared with a single dimer (Sec.~\ref{ssec:at}).

On proceeding towards the critical point (\ref{eq:Jcrit}), the maximum of 
the specific heat becomes progressively lower and narrower, developing a 
remarkably sharp profile close to $J_{\perp,c}$, as illustrated by the cases 
$J_\perp / J_\| = 1.42$ and $1.38$ in Fig.~\ref{fig:CompEDQMC}. We observe 
that the peak position falls by a factor of 3 from $J_\perp / J_\| = 2$ to 
1.42, while the gap, which is to the singlet branch of the two-triplon bound 
state, falls by a factor of 2.5 in units of $J_\|$ (Subsec.~\ref{ssec:spec} 
and Fig.~\ref{fig:LinfAllExec}). This change of effective temperature scale 
is apparent also in the rise of the susceptibility, which we characterize 
by the temperature, $T^\chi_{\rm half}$, where it has gained half of its peak 
height. In this case, $T^\chi_{\rm half}$ falls by a factor of 2.5 from 
$J_\perp / J_\| = 2$ to 1.42 while the triplet gap falls only by 30\%. 

\begin{figure}[t]
\includegraphics[width=\columnwidth]{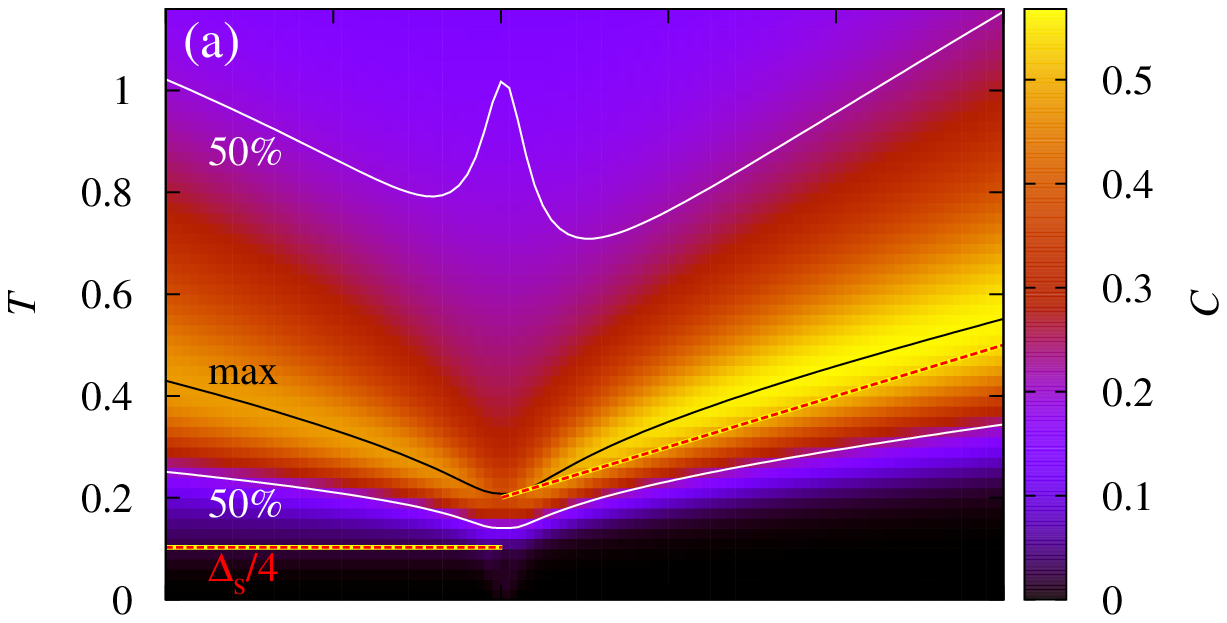}\\
\includegraphics[width=\columnwidth]{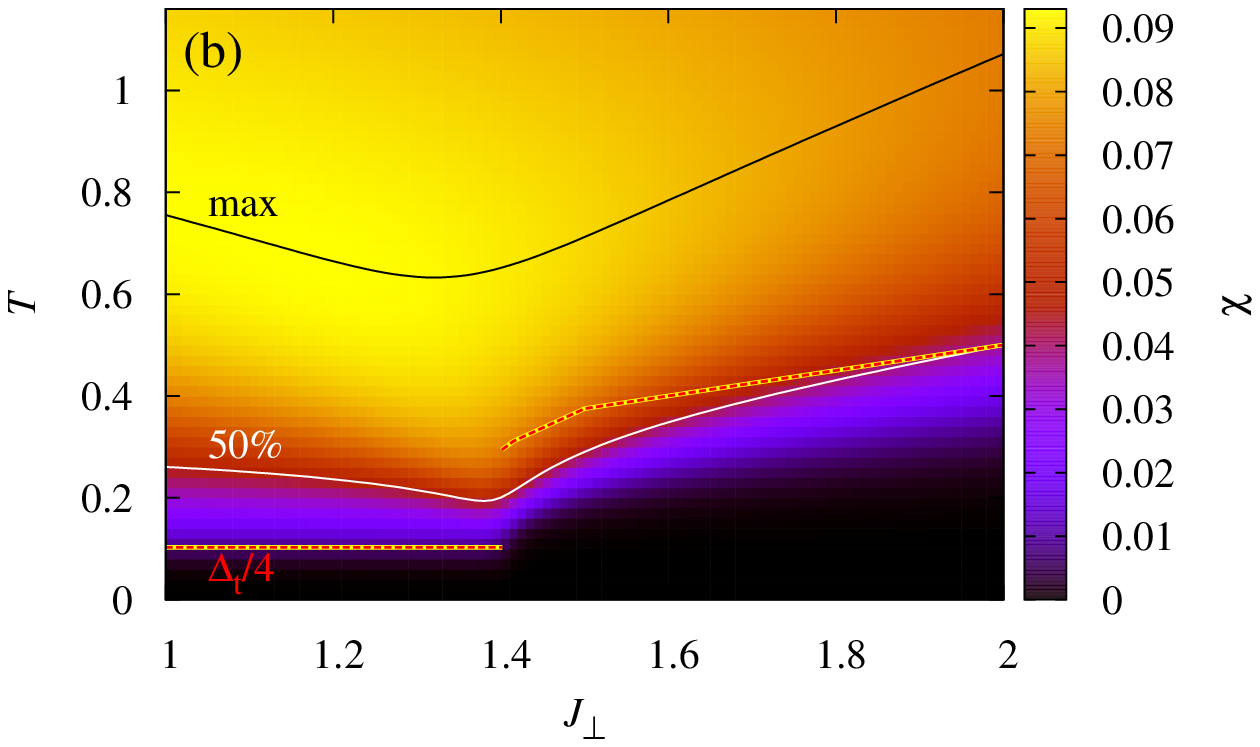}
\caption{(Color online) Exact diagonalization results for the specific heat, 
$C$ (a), and susceptibility, $\chi$ (b), of the fully frustrated ladder 
($J_\times = J_\| = 1$) with $L = 14$ rungs, shown for the full range of 
$J_\perp/J_\|$ ratios around the transition. Black lines mark the positions, 
$T^C_{\rm max}$ and $T^\chi_{\rm max}$, of the peaks in both quantities. White 
lines mark the temperatures, $T^C_{\rm half}$ and $T^\chi_{\rm half}$, where they 
have reached half of the peak height, and for $C(T)$ the temperature, 
$T^C_{\rm u}$, where the peak has fallen again to half of the peak height as 
$T$ continues to increase. Red-yellow dashed lines mark 1/4 of the gaps 
relevant for each quantity, which for $C(T)$ is the minimum gap $\Delta_s$ 
and for $\chi(T)$ the triplet gap $\Delta_t$ (Subsec.~\ref{ssec:spec}).}
\label{fig:ED14x2thermo}
\end{figure}

To study this evolution of the specific heat and susceptibility in more 
detail, Fig.~\ref{fig:ED14x2thermo} presents data for both quantities in 
contour form as a function of the coupling ratio, $1 \le J_\perp / J_\| \le 2$, 
and temperature, $T$. These results were obtained by exact diagonalization of 
a 14-rung ladder, where we recall (Subsec.~\ref{sec:FinSize}) that finite-size 
effects due to the intruder states (Fig.~\ref{fig:ED14x2exec}) occur 
near $J_{\perp,c}$. Indeed a very weak V-shaped feature is discernible in the 
specific heat at the lowest temperatures near $J_\perp = 1.4\,J_\perp$, but 
these effects remain so small for $L = 14$ that the qualitative behavior is 
not affected. The quantum phase transition (\ref{eq:Jcrit}) is clearly visible 
as a dip in both thermodynamic quantities as a function of $ J_\perp / J_\|$. 
This dip is more apparent, and more symmetrical, in the specific heat than in 
the susceptibility. The energy and temperature scales marked by the lines in 
Fig.~\ref{fig:ED14x2thermo} as an aid to characterizing the positions and 
widths of the peaks in $C$ and $\chi$ are discussed in detail in 
Sec.~\ref{sec:interp}. 

\begin{figure}[t]
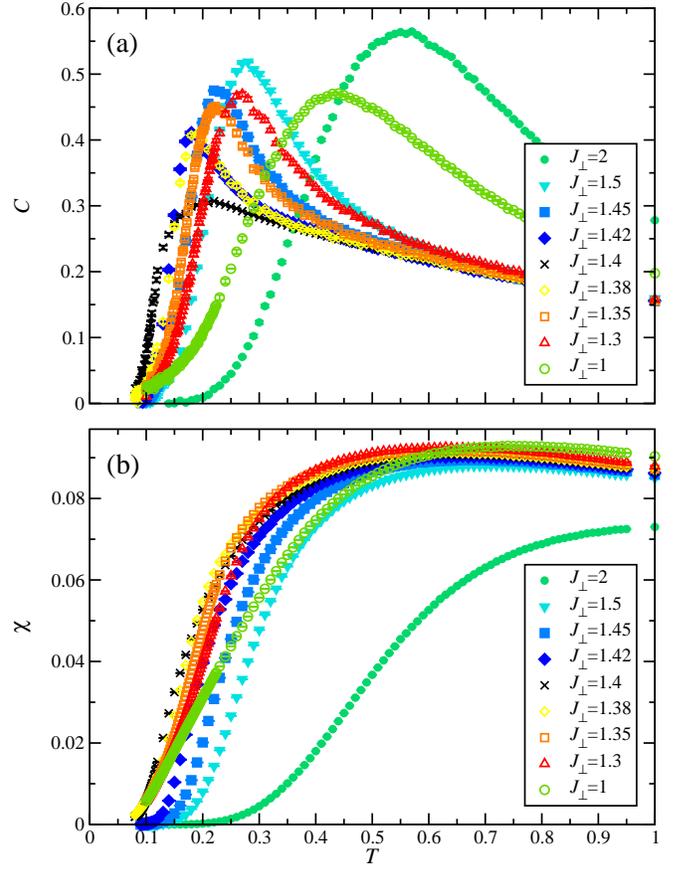

\includegraphics[width=\columnwidth]{Ccomp}\\[1mm]
\includegraphics[width=\columnwidth]{chiComp}
\caption{(Color online) QMC results for the specific heat, $C$ (a), and 
susceptibility, $\chi$ (b), of the fully frustrated ladder ($J_\times = J_\|
 = 1$) for a range of values of $J_\perp/J_\|$.}
\label{fig:QMC100x2thermo}
\end{figure}

For a fully quantitative investigation of the thermodynamic response in the 
regime around $J_{\perp,c}$, in Fig.~\ref{fig:QMC100x2thermo} we show QMC data 
for ladders of $L = 100$ rungs, focusing on the low-temperature region and 
adding the parameters $J_\perp / J_\| = 1.5$, 1.45, 1.4, 1.35, and 1.3 to those 
already shown in Fig.~\ref{fig:CompEDQMC}. Clearly the development (and 
disappearance) of the narrow specific-heat peak is very rapid, being 
concentrated largely into the region $1.3 \le J_\perp / J_\| \le 1.5$. The 
behavior of the specific heat is asymptotically completely symmetrical in 
the distance from the transition, as demonstrated by the fact that the 
results for $J_\perp / J_\| = 1.38$ and $1.42$ are difficult to distinguish. 
The case $J_\perp / J_\| = 1.4$ lies almost exactly at the transition, where 
we observe that the low-temperature peak structure is suppressed, leaving 
only a shoulder feature. Over the same range, the susceptibility exhibits 
rather little change in its broad maximum, $\chi_{\rm max}$, as a function 
of the ratio $J_\perp / J_\|$, but a quite dramatic drop in the position 
$T^\chi_{\rm half}$ of the half-height temperature, which falls by 40\% from 
$J_\perp / J_\| = 1.5$ to 1.4. As Fig.~\ref{fig:ED14x2thermo} also makes clear, 
the evolution of this quantity on the rung-triplet side of the transition is 
not symmetrical. Again we defer to Sec.~\ref{sec:interp} a discussion of how 
these effects arise in the absence of strong changes to the gap. 

We conclude this part of the presentation by returning to 
Figs.~\ref{fig:CompEDQMC}(d) and \ref{fig:CompEDQMC}(h), which represent 
a coupling ratio ($J_\perp = J_\|$) deep in the rung-triplet, or Haldane, 
phase of the ladder. Here we include results for the specific heat and 
susceptibility of a spin-1 Heisenberg chain \cite{YM93,Xiang98,FW05}, 
which were obtained by QMC simulations for chains of 60 sites and are 
taken from Ref.~\cite{cav2o4}. Because two $S = 1/2$ spins on a rung 
correspond to one spin $S = 1$, we have divided by twice the number of 
spins in the chain for a consistent normalization. The results for the 
$J_\perp = J_\|$ $(= J_\times)$ ladder and the spin-1 chain match completely 
in their low-temperature asymptotic behavior, reflecting the fact that the 
low-energy spectra of the two models are identical. However, the maxima of 
both $C$ and $\chi$ appear lower in temperature by a factor of 2 for the 
ladder than for the spin-1 chain, and the maximum of $C$ for the ladder is 
very much higher and sharper. Thus, it is clear that those parts of the 
ladder Hilbert space with rung spins $T_i = 0$ remain strongly relevant 
to the thermodynamic response at all but the lowest temperatures for 
parameters around $J_\perp = J_\|$, and we return to these contributions in 
Sec.~\ref{sec:interp}. As $J_\perp$ is decreased further, states including 
$T_i = 0$ rungs are pushed to successively higher energies until the spin-1 
Heisenberg chain is recovered from the ladder in the limit $J_\perp \to 
-\infty$.

\subsection{Frustration Effects}
\label{ssec:jtd}

\begin{figure}[t]
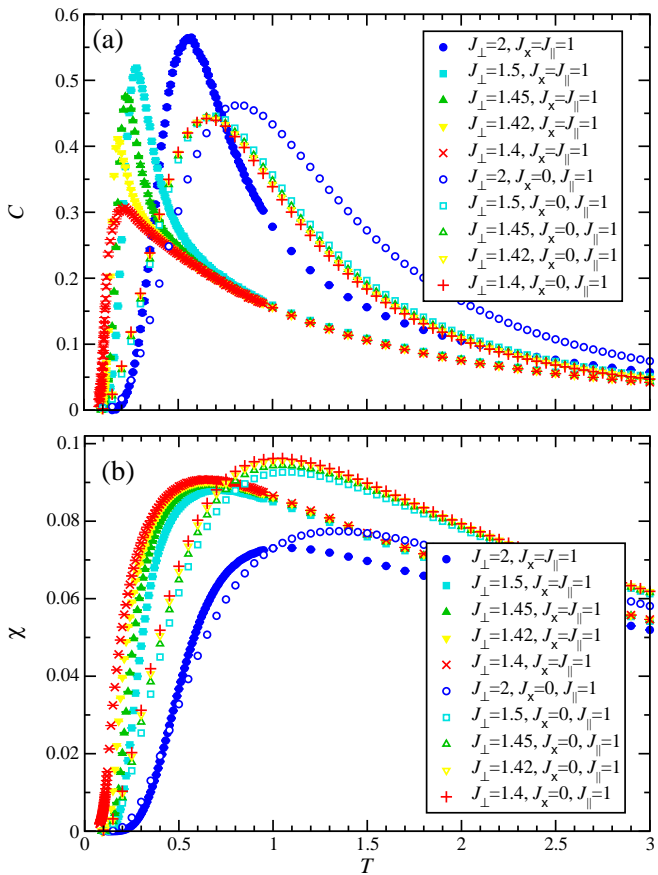

\includegraphics[width=\columnwidth]{CcompPL}\\[1mm]
\includegraphics[width=\columnwidth]{chiCompPL}
\caption{(Color online) Comparison between the specific heat, $C$ (a),
and susceptibility, $\chi$ (b), of a fully frustrated ladder ($J_\| = 
J_\times = 1$, closed symbols and crosses) and an unfrustrated ladder 
($J_\| = 1$, $J_\times = 0$, open and plus symbols) for a range of values of 
$J_\perp/J_\|$. Shown are QMC results for ladders of $L = 100$ rungs. Error 
bars on the unfrustrated ladder data are much smaller than the symbol sizes 
and are omitted.}
\label{fig:QMC100x2thermoPL}
\end{figure}

Following Subsec.~\ref{ssec:at}, we use our numerically exact QMC data 
also to assess the effects of frustration. For this we alter the diagonal 
(frustrating) coupling, $J_\times$ in Eq.~(\ref{eq:essh}), from 1 to 0 in units 
of $J_\|$, i.e.~we consider only the comparison with the unfrustrated ladder 
but, for reasons of space, avoid a systematic investigation of the crossover 
between these limits. The unfrustrated $S = 1/2$ two-leg ladder has been 
studied in considerable detail \cite{Frischmuth96,Weihong97,Gu99,Johnston00b},
including by QMC simulations to obtain the susceptibility \cite{Frischmuth96,
Johnston00b}. Here we have nevertheless generated our own data for both $C$ 
and $\chi$, using the SSE QMC technique \cite{Sandvik02}, and the comparison 
between the two types of ladder is shown in Fig.~\ref{fig:QMC100x2thermoPL}.
We restrict our attention again to the coupling region close to the phase 
transition (the discussion for weaker interdimer coupling may be found in 
Subsec.~\ref{ssec:at}) and focus only on the rung-singlet regime of the 
frustrated ladder, $J_\perp \ge 1.4\,J_\|$, as a comparison is not otherwise 
meaningful. 

The temperature unit in Fig.~\ref{fig:QMC100x2thermoPL} is the ladder leg 
coupling, $J_\|$, and thus the results offer a different perspective from 
that of Subsec.~\ref{ssec:at}, where $J_\perp$ was used. In these units, 
the specific-heat peak of the unfrustrated ladder moves slowly to lower 
temperatures as the coupling ratio is reduced. Still, the features of 
Fig.~\ref{fig:atc} remain clear at $J_\perp = 2\,J_\|$, in that the 
specific heat of the fully frustrated ladder has a significantly narrower 
and lower-lying maximum than that of the frustrated ladder. As the critical 
point  (\ref{eq:Jcrit}) is approached, the peak for the frustrated system 
shifts rapidly to lower temperatures, falling in height but becoming very 
much narrower. In the unfrustrated case, by contrast, the specific heat 
changes very little, as expected for such small variations of the parameters 
far from a phase transition, and thus stands as a constant benchmark of gap, 
peak, and broadening effects (Sec.~\ref{sec:interp}). 

The susceptibilities of the two cases show broadly similar qualitative 
features, with the positions ($T^\chi_{\rm max}$) of the maxima lying 
consistently lower in the frustrated than in the unfrustrated system. 
Dramatic shifts in temperature scales are less apparent in $T^\chi_{\rm max}$ 
than in the half-height temperatures, $T^\chi_{\rm half}$, with the curves for 
frustrated ladders moving strongly downward as $J_\perp = 1.4\,J_\|$ is 
approached, but no comparable behavior for unfrustrated ladders. The heights 
($\chi_{\rm max}$) of the $\chi$ peaks are quite consistent between the two 
cases, and notably lower than the decoupled-dimer result (\ref{eq:ChiDimer}) 
shown for comparison with $J_\perp = 2 \, J_\|$ in Fig.~\ref{fig:CompEDQMC}(e), 
indicating that the effects on $\chi_{\rm max}$ of spectral-weight redistribution 
away from a single energy $J_\perp$ are similar whether this occurs by the 
opening of a continuous triplon band or of many discrete bound states. 

We conclude that the anomalous behavior of both $C$ and $\chi$ in the fully 
frustrated ladder close to $J_{\perp,c}$ is indeed a signature of the many 
bound states emerging at low energies close to the frustration-induced 
first-order transition (Subsec.~\ref{ssec:spec}).

\section{Interpretation of Numerical Results}
\label{sec:interp}

In this section we compare the numerical results of Sec.~\ref{sec:results} 
with different analytical ones to gain further insight into the roles of 
frustration, of bound states, and of the large numbers of many-triplon bound 
states at low energies near the phase transition. We first consider the 
cluster description of Subsec.~\ref{ssec:at} to observe how much of the 
thermodynamic response is captured by short clusters of different lengths 
$n$. We then consider a large-$n$ description based on the statistical 
mechanics of unbound domain walls (between singlet and triplet rungs) to 
capture the unconventional effects of the high density of many-triplon 
excited states on the specific heat and susceptibility. We conclude by 
discussing the experimental consequences of and possibilities suggested 
by our conclusions, emphasizing that, in contrast to conventional gapped 
quantum magnets, $C(T)$ and $\chi(T)$ are characterized by different gaps
and by anomalous effective peak energy scales. 

\subsection{Small-Cluster Analysis}
\label{ssec:anc}

\begin{figure*}[p]
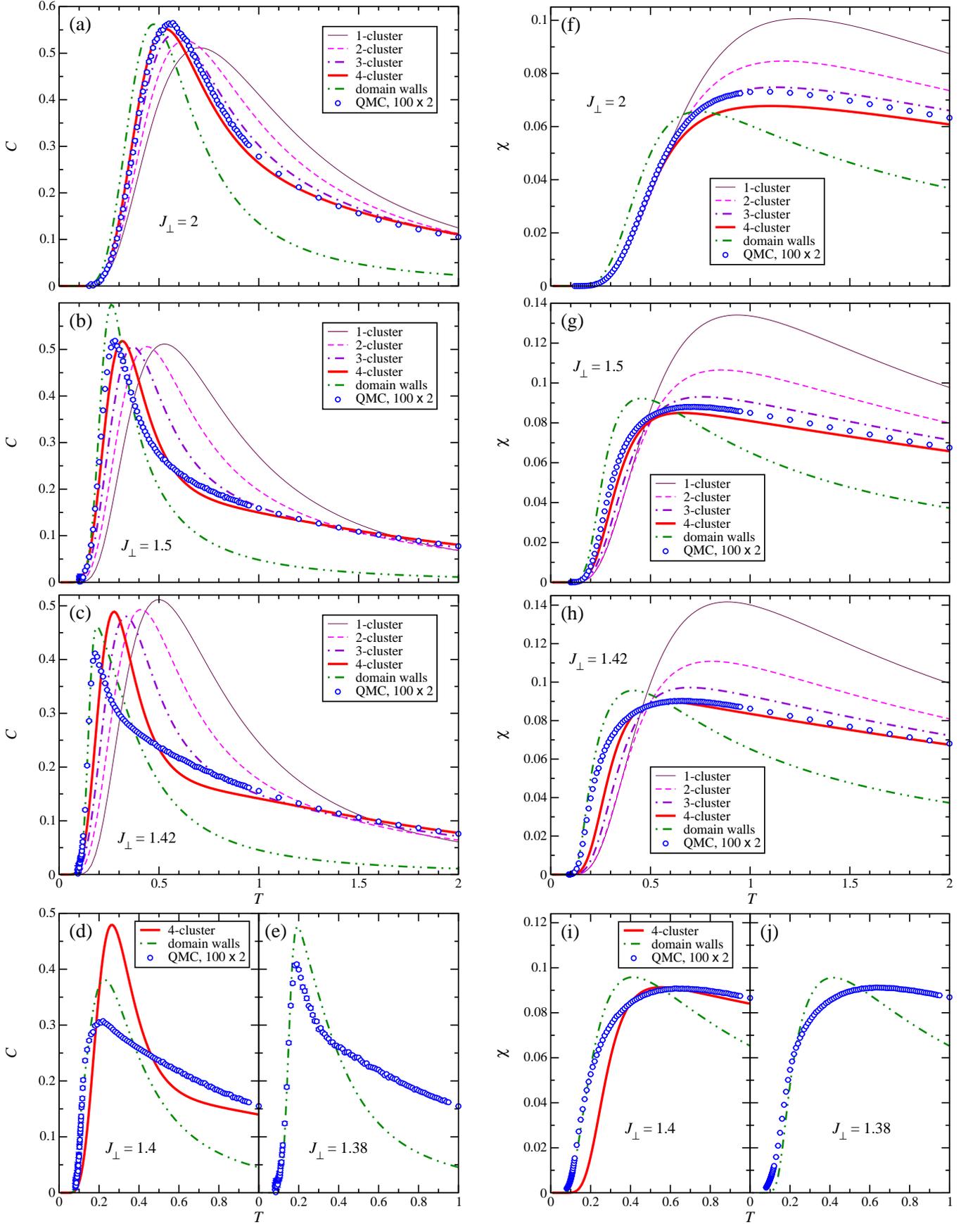

\includegraphics[width=\columnwidth]{clusterCJR2}\hfill%
\includegraphics[width=\columnwidth]{clusterChiJR2}\\
\includegraphics[width=\columnwidth]{clusterCJR1_5}\hfill%
\includegraphics[width=\columnwidth]{clusterChiJR1_5}\\
\includegraphics[width=\columnwidth]{clusterCJR1_42}\hfill%
\includegraphics[width=\columnwidth]{clusterChiJR1_42}\\
\includegraphics[width=\columnwidth]{clusterCJR1_4+1_38}\hfill%
\includegraphics[width=\columnwidth]{clusterChiJR1_4+1_38}
\caption{(Color online) Specific heat, $C$ (left column), and susceptibility, 
$\chi$ (right column), of fully frustrated ladders ($J_\times = J_\| = 1$) with 
rung coupling ratios $J_\perp/J_\| = 2$, 1.5, 1.42, 1.40, and 1.38. QMC results 
for ladders of $L = 100$ rungs are compared with approximate calculations 
based on clusters of 2, 3, and 4 rungs (Subsec.~\ref{ssec:at}) and on 
non-interacting domain walls (Subsec.~\ref{ssec:DW}).}
\label{fig:CompClusterQMC}
\end{figure*}

A preliminary interpretation of the numerical results presented in 
Sec.~\ref{sec:results} may be obtained by using the cluster approximation of 
Subsec.~\ref{ssec:at} to examine the extent to which the exact thermodynamic 
response functions are reproduced by considering only multi-triplon bound 
states up to a given size $n$. As in Sec.~\ref{sec:results}, we focus on the 
region around the phase transition, and of necessity on $J_\perp \ge J_{\perp,c}$ 
because the cluster approximation is applicable only in the rung-singlet 
regime. A comparison between the numerically exact results of 
Fig.~\ref{fig:QMC100x2thermo} and the analytical approximations of 
Subsec.~\ref{ssec:at} is presented in Fig.~\ref{fig:CompClusterQMC} for 
clusters of 2, 3, and 4 rungs and for four different values of $J_\perp/J_\|$. 
For reference we have also included a curve for the ``1-cluster,'' which is 
simply the response of the isolated dimer [Eqs.~(\ref{eq:Cdimer}) and 
(\ref{eq:ChiDimer})].

The cluster approximations for the specific heat 
[Fig.~\ref{fig:CompClusterQMC}] quantify the results in 
Figs.~\ref{fig:atc}(b) to \ref{fig:atc}(d), showing how longer clusters 
better capture the number of available low-lying states, thereby pushing 
the peak position, $T^C_{\rm max}$, systematically to lower values. The exact 
results demonstrate that the simple analytical approach is already accurate 
at the percent level for $J_\perp = 2\,J_\|$ when a cluster of 4 rungs is used 
[Fig.~\ref{fig:CompClusterQMC}(a)], and by extension it is more accurate 
still for all values $J_\perp > 2\,J_\|$. This result can be regarded as a 
consequence of the very short correlation length of the fully frustrated 
ladder for most values of $J_\perp$ away from the transition region. On 
closer inspection, the 4-rung approximation actually underestimates 
$T^C_{\rm max}$, while the 3-rung one overestimates it, illustrating the 
thermodynamic consequences of the low-lying energy levels shown in 
Fig.~\ref{fig:S1openEn} and Table \ref{tab:S1energies}.

Clearly at $J_\perp = 2\,J_\|$ the correction due to longer segments is 
actually towards energies above the lowest singlet of the 4-rung case. 
However, as $J_\perp$ is further reduced, the trend of the cluster 
approximations is primarily to change the peak heights, $C_{\rm max}$, rather 
than their positions, whereas the peak in the exact data continues to move 
to lower $T^C_{\rm max}$. This is a clear indication of the importance of the 
lowest levels in the spectra of ever-longer clusters moving below the 
one-triplon gap as $J_\perp$ approaches $J_{\perp,c}$. At $J_\perp = 1.5\,J_\|$ 
[Fig.~\ref{fig:CompClusterQMC}(b)], the 4-rung cluster is clearly no longer 
quantitatively adequate, and we comment that not only are extra bound states 
appearing at lower temperatures (around $0.3\,J_\|$) but also at higher 
ones over a range around $0.75 \,J_\|$. The cluster approximations change 
rather little in the critical regime [Figs.~\ref{fig:CompClusterQMC}(b) 
to \ref{fig:CompClusterQMC}(d)] and the results for $J_\perp = 1.42\,J_\|$ 
[Fig.~\ref{fig:CompClusterQMC}(c)] show clearly the effects of 
missing the very large numbers of low-lying states illustrated in 
Figs.~\ref{fig:ED14x2exec} and \ref{fig:LinfAllExec}, which cause the 
strong suppression of both $T^C_{\rm max}$ and $C_{\rm max}$. We draw attention 
again to the fact that these changes in the thermodynamic response are not 
brought about by changes in the gap, which is captured very well by the 
smallest cluster ($n = 2$, Subsec.~\ref{ssec:spec}); the differences at 
low temperatures visible below $J_\perp = 1.5\,J_\|$ are the consequence of 
increasing densities of states lying just above the gap. 

For the susceptibility, it is clear at $J_\perp = 2\,J_\|$ 
[Fig.~\ref{fig:CompClusterQMC}(f)] that all clusters provide an 
excellent description of $\chi(T)$ up to the half-height temperature, 
$T^\chi_{\rm half}$. Above this value, the dominant effect of longer clusters 
is to reduce the peak height, $\chi_{\rm max}$, and the 3- and 4-rung 
approximations begin to show a reduction in $T^\chi_{\rm max}$. In fact it is 
the 3-rung approximation that provides the best account of the exact data, 
with the 4-rung one causing too much suppression, which serves as a reminder 
that odd-$n$ clusters yield the lowest-lying triplets that govern $\chi(T)$. At 
$J_\perp = 1.5\,J_\|$ [Fig.~\ref{fig:CompClusterQMC}(g)], the gaps do begin to 
differ due to new lowest-lying triplet excitations (Subsec.~\ref{ssec:spec}), 
and neither the 3- nor the 4-rung approximation provides a good reproduction 
of the data, presumably reflecting the importance at this coupling ratio of the 
$n = 5$ triplet. At $J_\perp = 1.42\,J_\|$ [Fig.~\ref{fig:CompClusterQMC}(h)], 
the peak is by chance rather well described by the $n = 4$ cluster, but the 
situation at low $T$, where the $T^\chi_{\rm half}$ temperatures now differ 
strongly, illustrates the effects of large numbers of magnetic states 
whose origin lies in multi-triplon clusters with $n \ge 5$. 

In summary, comparisons between cluster approximations and the exact data 
show that only small clusters are perfectly sufficient to explain the response 
of the fully frustrated ladder over much of the rung-singlet regime. However, 
as the phase transition is approached, large numbers of bound states are moved 
below the one-triplon excitation energy and their effects are to reduce both 
the peak energy and the peak height in both $C$ and $\chi$. This demonstration 
of the importance of many-triplet excitations, and hence of the fact that 
large clusters ($n \gg 4$) must be considered when calculating thermodynamic 
properties, even when the system remains gapped, is the fundamental 
qualitative conclusion of the cluster analysis.

\subsection{Large-Cluster Analysis}
\label{ssec:DW}

With a view to capturing the rapid changes in thermodynamic response upon 
approaching $J_{\perp,c}$, and the underlying proliferation of low-lying 
excited states (Fig.~\ref{fig:LinfAllExec}), we consider a different type 
of picture based on the predominance of large-$n$ (many-triplon) clusters 
in the transition regime. This approximation takes clusters to be long and 
thus the domain walls separating rung singlets ($T_i = 0$) and triplets 
($T_i = 1$) to be sparse, such that they may be treated as non-interacting.
Although the full spectra of large-$n$ clusters are not known exactly, it 
is known (Subsec.~\ref{ssec:spec} and Fig.~\ref{fig:S1openEn}) that all 
many-triplon bound states have a low-lying level whose energy approaches the 
value $E_{\rm bond}$ [Eq.~(\ref{eq:S=1bond})]. A transfer-matrix formulation 
may be used to sum over the contributions from these levels of all large-$n$
clusters, providing a suitably size-extensive result expected to be at least 
qualitatively accurate in a regime where many-triplet bound states dominate 
the thermodynamic properties.

This treatment is based on the following specific assumptions. i) Each rung 
in the state $T_i = 0$ contributes $0$ to the total energy. ii) Each rung in 
the state $T_i = 1$ contributes $e_\infty + J_\perp$ to the total energy. iii) 
Each pair of domain walls contributes an energy $E_{\rm bond}$ to the total 
energy (each domain wall contributes $E_{\rm bond}/2$). iv) Each domain wall 
contributes a free spin 1/2, i.e.~a two-fold degeneracy. v) There are no 
interactions between the domain walls. In the parameter regime for which 
this analysis is designed ($J_\perp \rightarrow J_{\perp,c}$ from either side), 
the approximation is clearly not appropriate for small $n$, because the 
neglected binding effects are responsible for the low-energy levels visible 
on the left side of Fig.~\ref{fig:S1openEn}, but becomes systematically better 
for larger $n$, where many such states accumulate at energies around 
$E_{\rm bond}$. We comment that this treatment does allow for the occurrence 
of an odd number of domain walls, although these always appear in pairs on 
a ladder with periodic boundary conditions, and we expect this to be a 
legitimate approximation in the region close to the transition, where a 
finite concentration of domain walls is present.

The gas of non-interacting domain walls can be treated using the standard 
tools of statistical mechanics \cite{Ising25,HuangBook,BaxterBook}. The 
partition function of the gas of domain walls, whose energetic properties 
are specified as above, can be computed with the aid of the $2 \times 2$ 
transfer matrix
\begin{equation}
{\cal T} =
\begin{pmatrix}
1 & t (J_\perp,\beta,h) \\
t (J_\perp,\beta,h) & {\rm e}^{-\beta\, ( e_{\infty} + J_{\perp}) }
\end{pmatrix} \! ,
\label{eq:Tmat}
\end{equation}
where the first basis vector of the matrix corresponds to $T_i = 0$ and the 
second to $T_i = 1$, and
\begin{equation}
t (J_\perp,\beta,h) = 2\,{\rm e}^{-\beta\, ( e_{\infty}+ J_{\perp}+E_{\rm bond})/2 } \, 
\cosh (\beta\,h/2) \, .
\label{eq:TmatO}
\end{equation}
The arguments of the exponential functions in Eqs.~(\ref{eq:Tmat}) and
(\ref{eq:TmatO}) correspond to the bond energy between rungs $i$ and $i+1$ 
of the ladder, and thus the quantity $Z_L = {\rm Tr} {\cal T}^L$ is the sum
over the exponentials of the total energies of all such states for a ladder 
of $L$ rungs. Because the positions of the domain walls are arbitrary, the 
state degeneracies are accounted for automatically. The partition function 
at large $L$ converges to $Z_L \approx \lambda_+^L$, where $\lambda_+$ is 
the larger of the two eigenvalues, $\lambda_\pm$, of ${\cal T}$. The 
expression for $\lambda_+$ is lengthy and will not be presented here, but we 
comment that, in the high-temperature limit, $\lim_{\beta \to 0} \lambda_+ = 3$,
demonstrating that the domain-wall approximation retains on average three of 
the four states per rung. By analogy with Eqs.~(\ref{eq:cdef}) and 
(\ref{eq:chidef}), the specific heat and susceptibility per physical spin 
are given by
\begin{eqnarray}
C & = & \frac{\beta^2}{2} \, \left. \frac{\partial^2}{\partial \beta^2} \, 
\ln \lambda_+ \right|_{h=0} \, ,
\label{eq:Ctmat} \\
\chi & = & \frac{1}{2\,\beta} \, \left. \frac{\partial^2}{\partial h^2} \, 
\ln \lambda_+\right|_{h=0} \, 
\label{eq:chiTmat}
\end{eqnarray}
in the thermodynamic limit ($L \to \infty$). 

By inserting the values of $e_\infty$ [Eq.~(\ref{eq:Jcrit})] and $E_{\rm bond}$ 
[Eq.~(\ref{eq:S=1bond})], we calculate the specific heat and susceptibility 
for the values of $J_\perp/J_\|$ shown in Fig.~\ref{fig:CompClusterQMC}. We 
find that the domain-wall approximation reproduces the evolution of the 
low-temperature features near $J_{\perp,c}$ very much better than is possible 
using the short clusters of Subsec.~\ref{ssec:anc}. In both the specific heat 
[Figs.~\ref{fig:CompClusterQMC}(c) to \ref{fig:CompClusterQMC}(e)] and the 
susceptibility [Figs.~\ref{fig:CompClusterQMC}(h) to 
\ref{fig:CompClusterQMC}(j)] the exact response given by the QMC data is 
reproduced with remarkable accuracy up to a temperature of order $T_{\rm half}$. 
Given that the domain-wall approximation neglects both higher-lying excitations 
and binding effects, it is not surprising that its accuracy is restricted for 
high temperatures and far from the transition. However, results for the 
specific heat in Fig.~\ref{fig:CompClusterQMC} indicate that the domain-wall 
description provides an excellent estimate of $T^C_{\max}$, although it does
overestimate the number of states contributing to $C_{\rm max}$. Because the 
maximum of the susceptibility appears at higher temperatures, the domain-wall 
picture is less well suited to describe this feature. 

The qualitative and quantitative nature of our results confirms, most 
importantly, that at the transition a macroscopic number of excitations 
becomes relevant to the low-temperature thermodynamics of the fully frustrated 
ladder. Further, these states are well described by non-interacting domain 
walls between local segments of the ground states on both sides of the 
transition. We note that $J_{\perp,c} + e_{\infty} = 0$ at the transition, and 
therefore the only remaining energy scale is $E_{\rm bond}$, or more precisely 
$E_{\rm bond}/2 \approx 0.6\,J_\|$, which corresponds to the energy cost of a 
single domain wall. This explains the appearance of the low-$T$ maximum in 
$C(T)$ at $T \approx 0.2\,J_\|$ when $J_\perp \approx J_{\perp,c}$, where a 
factor of 3 between the intrinsic energy scale and the position of the 
maximum is typical (cf.~Fig.~\ref{fig:atc}). We also comment that, although 
the domain-wall energy ($E_{\rm bond}/2 \approx 0.6\,J_\|$) is larger than the 
Haldane gap ($\Delta \approx  0.4\,J_\|$), the domain wall is non-dispersive 
whereas the magnon of the spin-1 chain has a large bandwidth, and thus domain 
walls dominate the thermodynamics close to the transition even in the 
rung-triplet phase [Figs.~\ref{fig:CompClusterQMC}(e) and 
\ref{fig:CompClusterQMC}(j)].

\subsection{Characteristic Energy Scales of $C(T)$ and $\chi(T)$}
\label{ssec:aicchi}

One may seek to understand the form of the thermodynamic response by 
extracting the characteristic energy and temperature scales of $C(T)$ and 
$\chi(T)$. This conventional approach to the response of many low-dimensional 
models has been discussed in detail in Ref.~\cite{Johnston00b}. It may involve 
scaling the temperature axis by quantities such as $T_{\rm max}$ or the gap, 
$\Delta$, extracted from the low-temperature data, or the $y$-axes by the 
peak heights, $C_{\rm max}$ and $\chi_{\rm max}$, to seek similarities and 
differences between specific aspects of the datasets. 

In Fig.~\ref{fig:QMC100x2thermoPL}(a) we compared the specific heats of 
unfrustrated and frustrated ladders with different coupling ratios. We found 
that for unfrustrated ladders, where the (one-triplon) gap, $\Delta_t$, falls 
monotonically with decreasing coupling ratio due to both $J_\perp$ (band 
center) and $J_\|$ (band width), the peak position, $T^C_{\rm max}$, decreases 
only slowly in units of $J_\|$. By contrast, for frustrated ladders in the 
rung-singlet regime, the gap ($\Delta_t = J_\perp$) is locked to the coupling 
ratio for $J_\perp \ge 2$ and then becomes the singlet gap $\Delta_s = 2 \, 
(J_\perp - J_\|)$ of the two-triplon bound state as $J_\perp$ falls towards 
$J_{\perp,c}$, but $T^C_{\rm max}$ decreases significantly over this range. This 
contrast in behavior, in that the gap-to-peak ratio is much larger for 
frustrated ladders, is clearly visible in Figs.~\ref{fig:atc}(a) to 
\ref{fig:atc}(d). If the exponential factor $e^{-\Delta/T}$ is divided out 
of the low-temperature data, the remaining prefactor for unfrustrated ladders 
decreases when scaled to $J_\perp$, showing a decreasing density of low-lying 
states as the relative band width grows. For the frustrated ladders, the 
increasing prefactor shows the rising density of states close to, but just 
above, the gap as the quantum phase transition is approached 
(Subsec.~\ref{ssec:spec}). 

In the rung-triplet regime, the gap energy scale becomes the Haldane gap, 
which remains constant as $J_\perp$ is further reduced and appears quite 
unconnected to the increase in both $T^C_{\rm max}$ and $C_{\rm max}$ visible 
in Fig.~\ref{fig:QMC100x2thermo}(a). We have already observed that the 
Haldane chain does not give a good account of the specific-heat response, 
except in the very low-temperature limit [Fig.~\ref{fig:CompEDQMC}(d)]. As 
at $J_\perp > J_{\perp,c}$, the characteristic features of the specific heat 
for $J_\perp < J_{\perp,c}$ at intermediate temperatures are dominated by short 
clusters, this time of rung singlets in a background of rung triplets, which 
are responsible for the same high density of states near the transition. This 
conclusion is shown clearly both by the fact (Sec.~\ref{sec:results}) that 
the peaks for the rung-singlet and -triplet phases are almost completely 
symmetrical in their separation from $J_{\perp,c}$ and by the fact that the 
domain-wall analysis of Subsec.~\ref{ssec:DW} reproduces the high density 
of states on both sides of $J_{\perp,c}$.

Clearly the gap is not a particularly representative scale for the 
behavior of the specific heat, as may already have been anticipated from 
the red-yellow dashed lines in Fig.~\ref{fig:ED14x2thermo}(a), which for 
convenience of presentation show the normalized gap $\Delta/4$. Much more 
representative of the anomalous properties is $T^C_{\rm max}$, shown in the 
same figure by the black line, which captures the high density of bound states 
descending to low energies around the transition (Subsec.~\ref{ssec:DW}) and 
thus the corresponding dip in the response temperature. For this reason we do 
not attempt to scale our numerical data to $T^C_{\rm max}$ or $C_{\rm max}$, 
because the differences between parameter sets are compressed to very small 
regions at low temperatures. 

A further property of the specific heat is the width of its single peak. It 
is clear in Figs.~\ref{fig:CompEDQMC} and \ref{fig:QMC100x2thermo}(a) that 
the peak becomes very narrow as $J_\perp \rightarrow J_{\perp,c}$, but also that 
its height falls quite abruptly near $J_{\perp,c}$, leaving only a shoulder 
there [Fig.~\ref{fig:QMC100x2thermo}(a)]. In Fig.~\ref{fig:ED14x2thermo}(a), 
we marked with white lines the temperatures, $T^C_{\rm half}$ and $T^C_{\rm u}$, 
where the specific heat has risen to, and then fallen from, its maximum 
peak height as $T$ is increased. Taking their difference as a measure of 
the peak width, outside the transition region the width does indeed scale 
with $T^C_{\rm max}$, decreasing as the transition is approached. However, in 
the region $1.3 \le J_\perp \le 1.5$, $T^C_{\rm u}$ shows a strong spike, 
which makes the full-width at half height of the peak rise abruptly. 
This is a consequence of the rapid drop in peak height, visible in 
Fig.~\ref{fig:QMC100x2thermo}(a) but not reproduced well in either of our 
analytical approximations. We conclude that the narrowing of the peak is a 
generic property of the increasing density of low-lying states near the
quantum phase transition due to descending low-energy branches 
of large-$n$ bound states for all $n$; we suggest that the vanishing of 
the peak is a consequence of the reduction in relative weight of these 
branches compared to the very high multiplicity of states in each
bound-state multiplet.

Turning to the susceptibility, in Fig.~\ref{fig:QMC100x2thermoPL}(b) we 
compared its evolution with coupling ratio for unfrustrated and frustrated 
ladders. As for the specific heat, in unfrustrated ladders $T^\chi_{\rm max}$ 
decreases only weakly as the coupling ratio and the gap ($\Delta_t$) both 
fall. For frustrated ladders in the rung-singlet regime, the relative fall in 
peak position is much stronger, whereas the triplet gap is precisely $J_\perp$ 
over most of the phase diagram, dropping below this only when $J_\perp \le 1.5$, 
as detailed in Subsec.~\ref{ssec:spec}, and therefore has a very limited effect 
on the properties observed in Fig.~\ref{fig:QMC100x2thermoPL}(b). There is a 
small increase in $T^\chi_{\rm max}$ as $J_{\perp,c}$ is approached but, because 
the maximum of the susceptibility is so broad, a clearer indication of 
incipient critical behavior can be found from the rapid fall of the 
half-height temperature, $T^\chi_{\rm half}$, as $J_\perp \rightarrow 
J_{\perp,c}$. Once again, the low-temperature behavior of all curves in 
Fig.~\ref{fig:QMC100x2thermoPL}(b) is dictated only by the prefactor of the 
exponential, which reflects the rising density of triplets close to the gap 
energy as $J_\perp \rightarrow J_{\perp,c}$ (Subsec.~\ref{ssec:spec}). 

In the rung-triplet phase, where the lowest-lying (Haldane) mode is a 
triplet and the gap is a constant, the peak position of the susceptibility 
nevertheless moves to higher temperatures as $J_\perp$ decreases below 
$J_{\perp,c}$. Although in this case the rung-singlet and -triplet sides of 
the transition are not very symmetrical [Fig.~\ref{fig:ED14x2thermo}(b)], 
meaning that the relative effect of the Haldane mode is stronger, its 
contribution still does not account for much of the peak behavior, meaning 
away from the lowest temperatures [Fig.~\ref{fig:CompEDQMC}(h)], and this 
should again be ascribed to short clusters. 

Once again, the gap is not a very representative scale for the behavior of 
the susceptibility of the fully frustrated ladder, as shown by the red-yellow 
dashed lines in Fig.~\ref{fig:ED14x2thermo}(b) marking 25\% of the gap value. 
In this case, the quantity most representative of the anomalous properties 
around the transition appears to be $T^\chi_{\rm half}$, shown in the same 
figure by the white line. Once again, what is required is to capture the 
high density of triplet bound states descending to low energies in the 
vicinity of $J_{\perp,c}$ (Subsec.~\ref{ssec:DW}). In summary, the gaps 
change rather slowly across the critical region and are in general not the 
important quantities characterizing either the specific heat or the 
susceptibility of the fully frustrated ladder away from the low-$T$ limit. 
Instead, the falling peak positions are best captured by their temperature 
scales, which can be related to the low-lying many-triplon bound states of 
Subsec.~\ref{ssec:spec}. 

\subsection{Experimental Consequences}
\label{ssec:eccchi}

Despite the importance attached to the gap in low-temperature measurements of 
thermodynamic quantities in gapped quantum magnets, we have shown here that 
knowledge of the gap alone does not give much predictive power for the fully 
frustrated ladder. We draw attention again to the fact that the gaps 
extracted from the low-temperature behavior of the specific heat and of 
the susceptibility are not the same; although many frustrated systems are 
known in theory where low-lying singlet excitations may lie below the 
triplet gap, this remains very unconventional behavior from the standpoint 
of experimental observation. We reiterate that the specific heat is a measure 
of all states in the spectrum and in a frustrated system can be strongly 
affected by total-spin singlets (usually the lowest-lying states in an 
antiferromagnet) of multi-particle origin; this is very much the case in 
the fully frustrated ladder. By contrast, the susceptibility is a measure of 
finite-spin states only, and is usually dominated by total-spin triplets; 
multi-particle triplets are also particularly low-lying for odd-$n$ clusters 
in the fully frustrated ladder and the spin content of all excitations can 
be understood from the discussion of the spectrum in Subsec.~\ref{ssec:spec}.

In the context of the gap, a further valuable experimental quantity to discuss 
is the correlation length. This is not straightforward to define, but in 
principle should reflect the zero-range correlations between singlets and 
triplets on neighboring rungs. Because correlations are developed due to 
the presence of $n$-triplon bound states, the effective correlation length 
of the system should be an average over the effective cluster length and 
distribution. Although this average is expected to remain small, close to 
the phase transition there exist large but transient objects in the form 
of large-$n$ cluster bound states. We propose that $1/\Delta$ remains an 
appropriate measure of the effective correlation length, albeit with two 
different quantities, $1/\Delta_s$ and $1/\Delta_t$, required to characterize 
respectively non-magnetic and magnetic correlations near $J_{\perp,c}$ on the 
rung-singlet side.

Regarding the direct experimental relevance of our results, we noted in 
Sec.~\ref{sec:intro} that the material SrCu$_2$(BO$_3$)$_2$ shows a number 
of anomalous features in its thermodynamic response. The susceptibility 
has been found \cite{Kageyama99} to exhibit a sharp drop at a temperature 
about one tenth of the dominant coupling energy, whereas the specific heat 
\cite{Kageyama00a,Kageyama00b} has a sharp maximum near the same temperature. 
This system is thought to be a good realization of the (spin-1/2) 
Shastry-Sutherland model \cite{ShaSu81,MiUe03}, which is intrinsically 
two-dimensional, but shares two essential features of the fully frustrated 
ladder. One is that the one-triplon band is nearly flat, dispersing 
perhaps due primarily to higher-order Dzyaloshinskii-Moriya interactions. 
The other is that the magnetic interaction parameters deduced for this 
compound place it in a dimer-singlet phase, but very close to a quantum 
phase transition to a different ground state, thought to be a type of 
plaquette order \cite{Corboz13}. Certainly, the trends we observe for the 
thermodynamic properties of the fully frustrated ladder in the rung-singlet 
phase close to the transition are remarkably similar to SrCu$_2$(BO$_3$)$_2$, 
at least for the specific heat. The susceptibility of the ladder model is not 
as close (Fig.~\ref{fig:QMC100x2thermo}), with the temperature scale remaining 
higher and the maximum appearing more rounded. SrCu$_2$(BO$_3$)$_2$ is also 
known from two-magnon Raman \cite{Lemmens00,Gozar05} and inelastic neutron 
scattering measurements \cite{rglhcbdqc,rzrr} to show a highly anomalous
thermal evolution of the spectral weight, and this may also be interpreted 
\cite{HMN15} on the basis of the spectrum in Subsec.~\ref{ssec:spec}. We 
speculate that the type of behavior we have investigated here may arise in 
other frustrated spin systems with quantum phase transitions. 

In this context, it is worth considering the possibilities for the creation 
of non-thermal transitions in low-dimensional magnets. The most successful 
approach to date, applied in a number of systems, is the use of a hydrostatic 
pressure to alter the magnetic exchange interactions and thus to drive the 
system towards a critical point. In the event that this can be realized, 
strong changes can be expected in the spectrum that bring many levels to 
low energies or even to zero. Thus one would have the possibility of observing 
the type of evolution exhibited here for the fully frustrated ladder not 
only as a function of temperature but also as a function of pressure at 
a fixed, low temperature. 

Finally, we comment that another prime possibility for the creation of 
quantum phase transitions is the application of a magnetic field. The field 
couples only to magnetic states, and the effect would be to bring down one 
component of all the triplets. In a model as rich as the fully frustrated 
ladder, with significantly different singlet and triplet spectra, this would 
cause very strong and inequivalent alterations to $C$ and $\chi$. The phase 
diagram as a function of applied field and coupling ratio was considered in 
Ref.~\cite{rhmt} in the context of jumps and plateaus in the magnetization 
curve, whose nature has also been discussed in Ref.~\cite{lamas15}.
Deep in the rung-singlet regime (specifically, $J_\perp \gtrsim
1.58 J_\|$), the field-induced transition out of the rung-singlet phase
is expected to be first order, in fact to a mixed rung-singlet/-triplet
state with no threefold degeneracy on each rung and with a finite gap.
In the rung-triplet phase ($J_\perp < J_{\perp,c}$), the spectrum is
continuous beyond the one-magnon gap and therefore the applied field
causes a second-order transition to a gapless state. It was shown in
Ref.~\cite{rhmt} that there exists an intermediate regime, $J_{\perp,c} <
J_\perp \lesssim 1.58 J_\|$, where the field-induced transition is
first-order and also leads to a gapless state with triplets on all rungs.
The fully frustrated ladder in a magnetic 
field offers a further set of unconventional phenomena related to its exact 
bound states, flat bands, and high quasidegeneracies
of energy levels near the coupling-induced quantum phase transition. 

\section{Summary} 
\label{sec:summary}

We have investigated the thermodynamic properties of the fully frustrated 
two-leg spin-1/2 ladder. This system exhibits a first-order quantum phase 
transition between a rung-singlet regime for strong rung coupling and a 
rung-triplet, or Haldane, phase for weak rung coupling. For all values 
of the coupling ratio, the system has a gap to all excitations, and the 
magnetic specific heat and susceptibility are exponentially activated at 
low temperatures, followed by a single peak. However, in the vicinity of 
the transition point, the two quantities are characterized by different 
gaps in the low-energy spectrum on the rung-singlet side, and these gaps, 
the peaks, peak widths, and peak heights all show a dependence on the 
coupling constants quite different from reference systems such as the 
unfrustrated ladder, the spin-1 Heisenberg chain, or the frustrated $S = 1/2$ 
$J_1$--$J_2$ chain. 

The physics behind this anomalous behavior lies in the formation of 
multi-particle bound states. Single-rung excitations are localized objects, 
and so are their pairs, threesomes, and all higher $n$-excitation clusters, 
which may then be treated as open chains. On approaching the transition point, 
we observe the formation of large numbers of low-energy states, particularly 
singlets and triplets of the many-particle clusters, which lead to strongly 
enhanced fluctuations in this regime. These give rise to a sharp, and 
anomalously low-temperature, maximum in the specific heat and to an abrupt 
fall in effective response temperature of the susceptibility, results we 
show are quite different from the response of unfrustrated ladders. We 
comment that all of these effects may be expected to have a corresponding 
signature in the dynamical response function of the fully frustrated ladder 
at finite temperatures, which is the topic of a companion investigation 
\cite{HMN15}.

We have obtained extremely precise results by detailed numerical calculations 
using two techniques, ED and QMC, both of which work very well for this model. 
ED methods are usually restricted to small system sizes, which here we have 
extended to 28 sites by exploiting the exact relationship between 
the states of the fully frustrated ladder and of open and closed spin-1 
chains. The very short correlation length of the highly frustrated system is 
also advantageous for capturing most of the physics using a small system. The 
Hamiltonian in the rung basis (\ref{eq:exeh}) also ensures the complete 
absence of a sign problem in QMC simulations, allowing us to obtain results 
for ladders up to $L = 200$ rungs that are demonstrably far into the 
thermodynamic limit. Elimination of the sign problem by suitable choice of 
the basis is not new \cite{nakamura98}, but when applied in models where the 
total spin of the dimer (or other simplex unit) is conserved gives rise to 
the possibility of very powerful QMC approaches to frustrated spin systems. 
Methodological ideas and QMC 
results similar to ours for the fully frustrated ladder are reported in 
Ref.~\cite{radp}.
The next natural steps in this context are to test the performance of QMC 
simulations in the rung basis when the sign problem is not completely 
eliminated (for example in the ladder model of Eq.~(\ref{eq:essh}) with 
$J_\times \ne J_\|$ \cite{Weihong98,KSE99,Wang00,Starykh04,Kim08,HiSta10,Poilblanc10,
CCBZ15}) and to seek other geometries, interactions, and bases where the sign 
problem is absent. 

Two very general aspects of our results are of direct relevance to 
experimental studies of low-dimensional frustrated systems. One is the 
propensity of frustrated systems to bound-state formation, leading to the 
possibility of high densities of localized (narrow- or flat-band) excitations, 
which have a strong effect on the physical properties of the system. The other 
is the effect of proximity to a quantum phase transition, which leads to strong 
changes in the spectrum and may thus push high densities of states to low 
energies, with dramatic effects on both the thermodynamic and the dynamical 
response. Both features may already have been observed in the two-dimensional 
compound SrCu$_2$(BO$_3$)$_2$, a system whose ``Shastry-Sutherland'' geometry 
is so frustrated that the one-triplon band is almost completely flat and whose 
exchange constants are believed to place it in the dimer-singlet phase (the 
equivalent of the rung-singlet phase for the ladder), but very near the quantum 
phase transition to a suspected plaquette state. Indeed the specific heat, 
susceptibility, and dynamical structure factor measured for this material 
show some of the anomalous properties we have observed in one dimension in 
the fully frustrated ladder. The advances of this work, both in 
analytical understanding and in numerical capabilities, may be expected 
to assist in computing the thermodynamic properties of a broad range of 
frustrated models in the near future. 

{\em Note added in proof.} We regret to announce that one of our coauthors, Prof.\ 
Thomas Pruschke, passed away shortly after the acceptance of this 
article. We would like to express our gratitude for his unflagging 
support as a colleague and his incisive contributions as a physicist.

\acknowledgments

We thank H.\ R{\o}nnow for valuable discussions about SrCu$_2$(BO$_3$)$_2$, 
H.\ Tsunetsugu and W.-Q.\ Yu for helpful comments, and N.\ Chepiga for sharing 
with us some unpublished DMRG data for $S = 1$ chains. We are particularly 
grateful to F.\ Alet and K.\ Damle for discussions concerning the spin-dimer 
basis for QMC and for pointing out the feasibility of avoiding the QMC sign 
problem with additional $D^z D^z$ terms only.
We acknowledge the ALPS numerical libraries \cite{alps1,alps2}, with which 
preliminary QMC simulations were performed. This work was supported by the 
DFG research unit FOR1807, by the NSF of China under Grant 11174365, and by 
the National Basic Research Program of the Chinese MoST under Grant 
2012CB921704. SW acknowledges the hospitality of ECT${}^*$ (Trento) and the 
allocation of CPU time at JSC J\"ulich and RWTH Aachen through JARA-HPC. FM 
thanks the Paul Scherrer Institute for hospitality and the Swiss National 
Science Foundation for support.

\end{document}